\documentclass[iop]{emulateapj}

\submitted{Submitted to the Astrophysical Journal}

\def\ltapprox{\raise 2pt \hbox {$<$} \kern-1.1em \lower 5pt \hbox {$\approx$}}
\def\ltsim{\raise 2pt \hbox {$<$} \kern-1.1em \lower 4pt \hbox {$\sim$}}
\def\gtsim{\raise 2pt \hbox {$>$} \kern-1.1em \lower 4pt \hbox {$\sim$}}

\def\ie{{\it i.e.,~}}
\def\eg{{\it e.g.,~}}
\def\gtsim{\; \raise0.3ex\hbox{$>$\kern-0.75em \raise-1.1ex\hbox{$\sim$}}\; }
\def\ltsim{\; \raise0.3ex\hbox{$<$\kern-0.75em \raise-1.1ex\hbox{$\sim$}}\; }
\renewcommand{\thefootnote}{\alph{footnote}}	

\usepackage{multirow}
\usepackage{epstopdf}

\slugcomment{To appear in ApJ, 777, 141.}

\shorttitle{Revisiting scaling relations for giant radio halos in galaxy clusters}
\shortauthors{Cassano et al.}

\begin{document}


\title{Revisiting scaling relations for giant radio halos in galaxy clusters}


\author{R. Cassano \altaffilmark{1},
S. Ettori\altaffilmark{2,3}, 
G. Brunetti\altaffilmark{1},
S. Giacintucci\altaffilmark{4,5},
G.W. Pratt\altaffilmark{6},
T. Venturi\altaffilmark{1},
R. Kale\altaffilmark{1},
K. Dolag\altaffilmark{7,8},
M. Markevitch\altaffilmark{9}}

\altaffiltext{1}{INAF/IRA, via Gobetti 101, I--40129 Bologna, Italy}
\altaffiltext{2}{INAF/Osservatorio Astronomico di Bologna, via Ranzani 1, I--40127 Bologna, Italy}
\altaffiltext{3}{INFN, Sezione di Bologna, viale Berti Pichat 6/2, I-40127 Bologna, Italy}
\altaffiltext{4}{Department of Astronomy, University of Maryland, College Park, MD 20742-2421, USA} 
\altaffiltext{5}{Joint Space-Science Institute, University of Maryland, College Park, MD 20742-2421, USA}
\altaffiltext{6}{Laboratoire AIM, IRFU/Service dÕAstrophysique - CEA/DSM - CNRS - Universit\'e Paris Diderot, B\^at. 709, CEA-Saclay, 91191 Gif-sur-Yvette Cedex, France}
\altaffiltext{7}{University Observatory Munich, Scheinerstr. 1, D-81679 Munich, Germany}
\altaffiltext{8}{Max-Planck-Institut fŸr Astrophysik, Karl-Schwarzschild Strasse 1, Garching bei MŸnchen, Germany}
\altaffiltext{9}{Astrophysics Science Division, NASA/Goddard Space Flight Center, Greenbelt, MD 20771, USA}


\begin{abstract}

Many galaxy clusters host Megaparsec-scale radio halos, generated by ultrarelativistic electrons in the
magnetized intracluster medium. Correlations between the synchrotron power of radio halos and the
thermal properties of the hosting clusters were established in the last decade, including the connection
between the presence of a halo and cluster mergers. The X-ray luminosity and 
redshift limited Extended GMRT Radio Halo Survey provides a rich and unique dataset for statistical studies of
the halos. We uniformly analyze the radio and X-ray data for the GMRT cluster sample, and use the new 
Planck SZ catalog, to revisit the correlations between the power of radio halos and the thermal properties of galaxy
clusters. We find that the radio power at 1.4 GHz scales with the cluster X-ray (0.1--2.4 keV) luminosity 
computed within $R_{500}$ as $P_{1.4} \sim L^{2.1\pm0.2}_{500}$. Our bigger and more homogenous
sample confirms that the X-ray luminous ($L_{500} > 5\times 10^{44}$ erg\,s${^{-1}}$) clusters branch into 
two populations --- radio halos lie on the correlation, while clusters without radio halos have their radio upper limits well
below that correlation. This bimodality remains if we excise cool cores from the X-ray luminosities. 
We also find that $P_{1.4}$ scales with the cluster integrated SZ signal within $R_{500}$, measured by {\it Planck}, 
as $P_{1.4}\sim Y^{2.05\pm 0.28}_{500}$, in line with previous findings. 
However, contrary to previous studies that were limited by incompleteness and small sample size, we find that 
``SZ-luminous'' $Y_{500} > 6\times 10^{-5}$ Mpc$^2$ clusters show a bimodal behavior for the presence of radio halos, 
similar to that in the radio-X-ray diagram. Bimodality of both correlations can be traced to clusters
dynamics, with radio halos found exclusively in merging clusters. These results confirm the key role 
of mergers for the origin of giant radio halos, suggesting that they trigger the relativistic
particle acceleration.

\end{abstract}



\section{Introduction}

The presence of non-thermal components (relativistic particles and magnetic fields) mixed with the thermal Intra Cluster Medium (ICM) has 
been revealed by radio observations of galaxy clusters showing diffuse, giant Mpc-scale synchrotron {\it radio halos} (RHs) and {\it radio relics} in a substantial fraction of massive clusters (\eg Ferrari et al. 2008; Cassano 2009; Feretti et al. 2012 for reviews). 

Giant RHs are the most spectacular and best studied cluster-scale non-thermal sources.  Their origin is still poorly understood. One possibility is synchrotron emission from secondary electrons generated by the collisions between cosmic ray protons and thermal protons (so-called ``secondary models'', \eg Dennison 1980). However, the same collisions should produce gamma rays through the generation and decay of neutral pions. The non-detection of nearby galaxy clusters in $\gamma$-ray band at 0.1-100 GeV puts serious limits on the contribution of secondary electrons to the RH emission (Ackermann et al. 2010; Jeltema \& Profumo 2011; Brunetti et al. 2012).

A second possibility is that the turbulence, generated in the ICM during cluster-cluster mergers, re-accelerates preexisting GeV electrons (\eg Brunetti et al. 2001; Petrosian 2001). The ``historical'' motivation for turbulent acceleration for the origin of RHs comes from the high-frequency steepening of the Coma halo spectrum, implying that the mechanism responsible for the acceleration of the emitting electrons is inefficient (\eg Schlickeiser et al. 1987). More recently, the discovery of RHs with extremely steep spectra\footnote{Here we adopt the convention $f_{\nu}\propto\nu^{-\alpha}$.}, $\alpha\sim1.5-2$, support turbulent re-acceleration and disfavor a ``secondary'' origin of giant RHs (\eg Brunetti et al. 2008; Dallacasa et al. 2009; Giovannini et al. 2009, Macario et al. 2010, 2011; Giacintucci et al. 2011, 2013; Bonafede et al. 2012; Venturi et al. 2013). 

Studies of statistical properties of giant RHs in clusters and their connection with the cluster dynamics are extremely useful to constrain the origin of halos. 
From the inspection of the NRAO VLA Sky Survey (NVSS, Condon et al. 1998) fields containing X-ray selected galaxy clusters, Giovannini et al. (1999) concluded that RHs are rare at low X-ray luminosities ($L_X\ltsim 10^{45}\,h_{50}^{-2}$ erg/sec), while only the most X-ray luminous systems host RHs, with a probability of $\sim 1/3$. Since then a number of correlations have been found between thermal and non-thermal cluster properties, suggesting a tight connection between them. In particular, the synchrotron monochromatic radio power of halos at 1.4 GHz ($P_{1.4}$) has been found to increase with the cluster X-ray luminosity, temperature and total mass (\eg Colafrancesco 1999; Liang 1999; Feretti 2002; Govoni et al. 2001; En\ss lin \& R\"ottgering 2002; Feretti 2003; Cassano et al. 06). These scalings call into question the rarity of halos in clusters of low X-ray luminosity, suggesting that the lack of RH detections in those clusters may result from the combination of the radio power--X-ray luminosity correlation and the sensitivity of the radio survey (\eg Kempner \& Sarazin 2001).

\noindent There is also substantial evidence that RHs are found in clusters with significant substructure in the X-ray images, as well as complex gas temperature distribution, which are signatures of cluster mergers (\eg Schuecker et al. 2001; Govoni et al. 2004; Markevitch \& Vikhlinin 2001). In particular, Buote (2001) provided the first quantitative comparison of the dynamical state of clusters with RH, discovering a correlation between the RH luminosity at 1.4 GHz and the magnitude of the dipole power ratio $P_1/P_0$, which is a measure of the cluster's X-ray morphological disturbance. However, these observational claims were based on collections of data from the literature and not on statistical samples of galaxy clusters.

An important step was recently obtained through deep radio observations of a complete sample of galaxy clusters as part of the Giant Metrewave Radio Telescope ({\it GMRT}) RH Survey (Venturi et al. 2007; 2008; GRHS hereafter). These observations confirmed that RH are not ubiquitous in clusters. They are found only in $\sim$30\% of the X-ray luminous systems ($L_X(0.1-2.4\,\mathrm{keV})\geq5\times10^{44}$ erg/s). The sensitivity reached by these observations allowed for the first time to place deep upper limits on the diffuse radio flux of clusters without giant RH and to show that clusters branch into two populations: RHs trace the correlation between $P_{1.4}$ and $L_X$, while the upper limits on the radio luminosity of clusters with no RH lie about one order of magnitude below that correlation (\eg Brunetti et al. 2007, 2009). Using several methods to characterize cluster substructures, it was also shown that clusters with and without RH can be quantitatively differentiated in terms of their dynamical properties, with RHs always associated with dynamically disturbed clusters while clusters without RHs being more ``relaxed'' (Cassano et al. 2010).

\noindent Sensitivity is critical in these studies. Indeed analyses based on all-sky surveys, such as the NVSS and WENSS that have 
a sensitivity $4-5$ times worse than the GRHS, do not allow to recover a bimodal behavior in the radio-X--ray diagram (\eg Rudnick \& Lemmerman 2009 for the WENSS). On the other hand, evidences for a bimodal behavior of clusters was recently found through a stacking analysis of clusters in the SUMSS (Brown et al. 2011). 

More recently, Basu (2012) cross-correlated the {\it Planck} ESZ cluster catalogue (Planck Collaboration 2011a) with radio data from the GRHS. He found a correlation between $P_{1.4}$ and the integrated Sunyaev-Zel'dovich (SZ) effect measurements, but did not find strong indication for a bimodal split between RH and radio-quiet clusters. To explain this apparent lack of bimodality in SZ, Basu (2012) suggested that X-ray observations could be biased towards the detection of low-mass cool-core clusters, whereas SZ selection picks up the most massive systems, irrespective of their dynamical states. 

In this paper, we improve on the previous statistical studies on the distribution of clusters in the $P_{1.4}$-$L_X$ diagram by using RH and clusters with radio upper limits from the GRHS and its extension and, when needed, including RHs from the literature. Contrary to previous analyses that used cluster X-ray and radio halo luminosities from the literature, we reevaluate the radio and X-ray luminosities in a homogeneous way. In particular, we derive the X-ray luminosity within $R_{500}\footnote{$R_{500}$ is the radius corresponding to a total density contrast 500$\rho_c(z)$, where $\rho_c(z)$ is the critical density of the Universe at the cluster redshift.}$ and include the correction due to the contribution of the cool core (when present). Furthermore, since the integrated SZ signal is a more robust indicator of the cluster mass than the X-ray luminosity (\eg Motl et al. 2005; Nagai 2006), we cross-checked our sample with the recent {\it Planck} SZ cluster catalogue (PSZ; Planck Collaboration et al. 2013b) and derived the distribution of clusters in the radio-SZ plane. 

In Sect.~2, we describe the cluster sample; in Sect.~3, we summarize the procedure to derive different cluster quantities (radio-halo power; X-ray luminosity, SZ flux, morphological parameters), identify cool-core clusters in the sample, and fit the scaling relations. In Sect.~4, we report on the expected theoretical scalings. We derive the distribution of clusters in the radio--X-ray diagrams in Sect.5, and in the radio-SZ (mass) diagrams, in Sect.~6. Finally, in Sect.~7, we give summary and conclusions.

A $\Lambda$CDM cosmology ($H_{o}=70\,\rm km\,\rm s^{-1}\,\rm Mpc^{-1}$, $\Omega_{m}=0.3$, $\Omega_{\Lambda}=0.7$) is adopted.

\section{The sample}
\label{sec:sample}

The GRHS is a deep, pointed radio survey of clusters selected from the {\it ROSAT}--ESO Flux Limited X--ray (REFLEX; B\"oringher et al. 2004) and extended {\it ROSAT} Brightest Cluster Sample (eBCS; Ebeling et al. 1998, 2000) catalogues. These two catalogues have almost the same flux limit in the $0.1-2.4$ keV band ($\gtsim 3\cdot 10^{-12} \rm{erg\,s^{-1}\,cm^{-2}}$) and their combination yields a homogeneous, flux-limited sample of clusters.
The GRHS consists of 50 galaxy clusters with z=$0.2-0.4$, X-ray luminosity $L_X>5\times 10^{44}$ erg/s and declination $\delta\geq 30^{\circ}$ for the REFLEX sample and $15^{\circ}\leq\delta\leq60^{\circ}$ for the eBCS sample. 
With the above selection criteria, the sample is X-ray luminosity-limited up to $z\simeq 0.25$ and X-ray flux-limited at higher redshift\footnote{This implies a minimum $L_X\sim 10^{45}$ erg/s 
at $z\sim0.35$.}(see Fig. 1 and 2 in Cassano et al. 2008).  

\noindent Recently, we have undertaken an extension of the GRHS by considering all clusters in the REFLEX and eBCs catalogs with $\delta> -30^{\circ}$ with the same $z$ and $L_X$ selection (Kale et al. 2013). This extension yields a final sample of 67 galaxy clusters, which we refer to as the extended {\it GMRT} RH Survey (EGRHS). 
For all clusters in the EGRHS with the radio data already available, we searched the {\it ROSAT} and {\it Chandra} archive and found data for a sub-sample of 40 galaxy clusters: 

\begin{itemize}
\item[-] 29 with radio upper limits 
\item[-] 8 with giant radio halos
\item[-] 3 with radio halos with ultra-steep spectra\footnote{We do not include a forth one, RXCJ1514.9-1523 (Giacintucci et al. 2011), which was only recently observed by {\it Chandra}, the data analysis is still ongoing (Giacintucci et al. in prep.).}.

\end{itemize}

In addition to clusters belonging to the EGRHS, we also searched in the X-ray archive and found data for 14 clusters with RHs from the literature:

\begin{itemize}
\item[-] 11 with giant radio halos
\item[-] 3 with radio halos with ultra-steep spectra
\end{itemize}

\noindent The total sample with radio and X-ray information consists of 54 galaxy clusters, whose main properties are reported in Table~\ref{Tab.ULRH}:

\begin{itemize}
\item[-] 8 RH from the EGRHS 
\item[-] 29 clusters with radio upper limits from the EGRHS
\item[-] 6 RH with ultra-steep spectra ($\alpha>1.5$; USSRH hereafter). Not to be compared to the upper limits (which were scaled at 1.4 GHz with $\alpha=1.3$). 
\item[-] 2 RH from the literature, Abell 1995 and the Bullet cluster, which are in the same redshift and X-ray luminosity range of the EGRHS 
\item[-] 9 RH from the literature, which do not fulfill the EGRHS selection criteria (in redshift and X-ray luminosity)   
\end{itemize}

\noindent The inclusion of RH from the literature is useful to have a sufficient leverage in radio/X-ray luminosities that may help to derive the scaling relations, however the comparison between halos and upper limits will be performed only for clusters of EGRHS, which are in the same redshift range.

\section{Data analysis}

In this Section we briefly describe the procedures undertaken to derive the radio and X-ray luminosities of clusters, to identify
cool core clusters and to analyze the cluster dynamical status. We also report measurements of the Sunyaev-Zel'dovich signal
found in the literature.

\subsection{Radio power of halos}
\label{Sect.radio}

Table 1 reports the radio halo powers and upper limits for the clusters
in the sample. We refer to the literature information (see notes to the table) for details and for the radio images.

For 12 RH clusters in the sample we re--analysed archival 1.4 GHz VLA--C and
VLA--D  array data (Giacintucci et al. in prep.). Other 6 clusters with giant RHs are published by our group 
and the data analysis was carried out following procedures similar  to
those described in this Section. For the remaining 
7 RHs the radio flux densities were taken from the literature.
For those clusters analyzed by us (12+6), the flux density 
of the radio halo was measured from low angular resolution images obtained  
after subtraction of the contribution of the individual radio sources 
embedded in the diffuse emission. 
In particular, we identified the discrete radio galaxies in (or projected 
onto) and around the cluster region using the higher resolution images
produced with the C--array datasets (when available).
The discrete radio sources were subtracted 
from the u--v datasets, and the resulting u--v visibilities were
then used to image the radio halo emission at low resolution.
In those cases where high resolution data were not available, to evaluate
and subtract the contribution of individual sources we produced images 
gapping the innermost region of the u--v plane and using only the remaining 
long baselines ($\gtsim 1-2 \lambda$), that contain information on 
structures on angular scales smaller than the underlying large--scale radio 
halo. For extended sources, we used sets of images with different resolutions 
and/or u--v ranges to determine their total extent and morphology. 
For each source and for each cluster, we carefully checked that the total 
flux density subtracted from the u--v data is consistent with the flux density 
measured on the images. 

We measured the total flux density of the radio halos starting from the 
3$\sigma$  countour level in the final images, then we progressively 
increased the extraction region until the integrated flux density reached a 
maximum value, and considered this maximum value as the total flux density of 
the halo. 
This procedure leads to an average increase of the halo flux density by only  
$\sim$ 5\% with respect to the value within the 3$\sigma$
isocontour.

Finally, we calculated the corresponding radio power at the cluster 
redshift and applied a k--correction $(1+z)^{(1-\alpha)}$, where the spectral 
index is taken from the literature (references in Tab.1), or it is assumed  to be $\alpha=1.3$ when not available. 
The errors on the diffuse radio flux density $f_H$ account for the uncertainty in the calibration of 
the absolute flux density scale, the error due to the noise in the 
integration area and the error due to the subtraction of the discrete radio 
sources in the halo region, as:

\begin{equation}
\sigma_{f_H}=\sqrt{(\delta_{cal}\,f_H)^2+(rms\,\sqrt{N_{beam}})^2+\sigma_{sub}^2}
\end{equation}

\noindent where $\delta_{cal}$ is typically of the order of 5-8\%, $rms$ is the noise of the map,  $N_{beam}$ is the number of independent beam in the halo region, and $\sigma_{sub}$ is the error due to the uncertainty in the source subtraction.
$\sigma_{f_H}$ does not account for the uncertainty due to the missing short spacings in the u--v coverage of the interferometric observations,
and this may bias the flux densities towards lower values.

Upper limits to the diffuse radio flux of clusters without giant RH were reported in Venturi et al. (2008) and Kale et al. (2013) and scaled at 1.4 GHz using a spectral index typical of RHs, $\alpha=1.3$.

\subsection{X-ray luminosities}
\label{Sect.Lx}

To derive the cluster X-ray luminosities we use {\it ROSAT} data, PSPC preferentially and HRI when PSPC data are not available. In those cases where {\it ROSAT} data are not available we use {\it Chandra} data (see Table~\ref{Tab.ULRH}). For all clusters we derive the X-ray luminosities inside $R_{500}$ centered on the centroid of the X-ray emission. To estimate $R_{500}$ for our clusters we searched in the literature for information about the X-ray temperature and then applied the relations
from Arnaud et al. (2005). 
We derived luminosities in the $0.1-2.4$ keV band in three different ways: {\it i)} the entire cluster emission inside $R_{500}$,
denoted as $L_{500}$; {\it ii)} the emission inside the aperture $[0.15-1]R_{500}$, denoted as $L_{500,nc}$; {\it iii)} $L_{500,cor}$, the X-ray luminosity inside $R_{500}$ corrected for the excess emission within $0.15 R_{500}$, due to the presence of a cooling core.
For each cluster, $L_{500,cor}$ is computed by performing a fit with a $\beta$-model to the cluster emission outside $0.15 R_{500}$, fixing $r_c=0.15R_{500}$ (assuming that $r_c\simeq r_{cool}\simeq 0.15 R_{500}$, which is $\sim 100-200$ kpc for our sample) and evaluating the contribution of the model inside $0.15 R_{500}$. When the model fit underestimates the counts in the core we correct the central region by using the fit to the X-ray brightness distribution outside $0.15 R_{500}$. 
 We masked the detected point sources after a careful inspection of the cluster {\it Chandra} images (only in four cases we used {\it ROSAT} PSPC/HRI data). Following Russell et al. (2013), bright central point sources were identified and masked using the {\it Chandra} 5-8 keV images.The flux in the masked regions has been replaced by estimates based on the cluster best-fit model for the spatial brightness distribution. The obtained value of $L_{500}$ and $L_{500,cor}$ are reported in Table~\ref{Tab.ULRH}.

%
%

\begin{deluxetable*}{lcccccclcc}
\tabletypesize{\scriptsize} 
\tablecolumns{10} 
\tablewidth{514.48149pt}
\tablecaption{Cluster's properties\label{Tab.ULRH}}
\tablehead{
\colhead{cluster name} & \colhead{RA$_{J2000}$} &  \colhead{DEC$_{J2000}$} & \colhead{z} &  \colhead{$L_{500}$} & \colhead{$L_{500,cor}$}  &\colhead{$L_{core}/L_{500}$} & \colhead{$P_{1.4}$} & \colhead{X-ray} & \colhead{SZ}}
\startdata
\sidehead{Upper limits (EGRHS) }
A2697 				     &  00 03 11.8 &  $-$06 05 10 & 0.232 &   7.29$\pm$0.41 &  7.29$\pm$0.41  & 0.34 & $<$0.41$^v$ & 19 (H)& $\surd$\\
A141  				     &  01 05 34.8 &  $-$24 39 17 & 0.230 &   6.82$\pm$ 0.27 & 6.82$\pm$ 0.27 &  0.13& $<$0.43$^v$ & 33 (H)& $\surd$\\ 
A3088 				     &  03 07 04.1 &  $-$28 40 14 & 0.254 &   6.97$\pm$0.09  & 5.63$\pm$0.08 & 0.46 & $<$0.43$^v$ & 19 (C)& $\surd$\\
RXCJ0437.1$+$0043 &  04 37 10.1 &  $+$00 43 38 & 0.284 &   6.99$\pm$0.08  & 6.15$\pm$0.08 & 0.45 &$<$0.65$^k$ & 30 (C)\\
RXCJ1115.8$+$0129 &  11 15 54.0 &  $+$01 29 44 & 0.350 &  12.69$\pm$0.11 & 8.21$\pm$0.10 & 0.63$^{cc}$ & $<$0.47$^v$ & 39 (C)& $\surd$\\
A2485 			   	     &  22 48 32.9 &  $-$16 06 23 & 0.247 &   3.27$\pm$0.07  & 3.07$\pm$0.07 & 0.39& $<$0.47$^k$ & 20 (C)\\ 
A2631 				     &  23 37 40.6 &  $+$00 16 36 & 0.278 &   8.62$\pm$0.70 & 8.62$\pm$0.70 & 0.21 & $<$0.41$^v$&15 (H)& $\surd$\\
A2645 				     &  23 41 16.8 &  $-$09 01 39 & 0.251 &   4.13$\pm$0.4 & 4.13$\pm$0.4 & 0.43 &$<$0.59$^q$& 35 (H)& $\surd$\\ 
A2667 				     &  23 51 40.7 &  $-$26 05 01 & 0.226 &  12.50$\pm$0.4 & 10.94$\pm$0.4 &0.46$^{cc}$ &  $<$0.45$^v$ & 21 (H)& $\surd$\\
Z348                               &  01 06 50.3 &  $+$01 03 17 & 0.255 & 6.30$\pm$0.60  & 4.26$\pm$0.60  & 0.54$^{cc}$  &$<$0.65$^k$ &13 (H)\\
RXJ0142.0$+$2131     &  01 42 03.1 &  $+$21 30 39 & 0.280 &  6.00$\pm$0.10  & 6.00$\pm$0.10 & 0.34 &  $<$0.45$^k$ & 20 (C)& $\surd$\\
A267	                           &  01 52 52.2 &  $+$01 02 46 & 0.230   & 6.29$\pm$0.44 &  5.94$\pm$0.44  &   0.36 &  $<$0.34$^k$& 16 (H)& $\surd$\\
RXJ0439.0$+$0715     &  04 39 01.2 &  $+$07 15 36 & 0.244 & 8.05$\pm$0.59 & 7.69$\pm$0.58 &    0.36  & $<$0.46$^k$ &19 (H)& $\surd$\\
RXJ0439.0$+$0520     &  04 39 02.2 & $+$05 20 43  & 0.208  & 5.35$\pm$0.47  & 4.05$\pm$0.46 &    0.55$^{cc}$   & $<$0.32$^k$& 12 (H)\\   
A611                                &  08 00 58.1 &  $+$36 04 41 & 0.288 & 4.96$\pm$0.64 & 4.96$\pm$0.64 & 0.46 & $<$0.43$^v$ & 17 (H)& $\surd$\\
Z2089                              &  09 00 45.9 &  $+$20 55 13 & 0.235 & 5.28$\pm$0.41 & 3.98$\pm$0.41 & 0.52$^{cc}$ & $<$0.26$^v$& 17 (H)\\
A781                                &  09 20 23.2 &  $+$30 26 15 & 0.298 & 5.44$\pm$0.14 & 5.44$\pm$0.14 & 0.12 & $<$0.36$^v$ & 10 (C)& $\surd$\\
Z2701                              &  09 52 55.3 &  $+$51 52 52 & 0.214 &   4.72$\pm$0.43 & 3.38$\pm$0.42 & 0.55$^{cc}$  &  $<$0.35$^v$ & 10 (H)\\
A1423                              &  11 57 22.5 &  $+$33 39 18 & 0.213 &  5.35$\pm$0.37  & 4.76$\pm$0.38  &0.41&   $<$0.38$^v$ & 19 (H)& $\surd$\\
A1576                              &  12 36 49.1 &  $+$63 11 30 & 0.30  &   6.68$\pm$0.14 & 6.38$\pm$0.14 & 0.24& $<$0.64$^k$ &17 (P)& $\surd$\\
RXJ1532.9+3021          &  15 32 54.2 &  $+$30 21 11 & 0.345 &  17.94$\pm$0.27  & 12.28$\pm$0.22 & 0.64$^{cc}$& $<$0.66$^v$ & 22 (C)\\
A2146                             &  15 56 04.7 &  $+$66 20 24 & 0.234 &  5.69$\pm$0.04  & 5.69$\pm$0.04 & 0.42 & $<$0.39$^r$ & 65 (C)& $\surd$\\
A2261                             &  17 22 28.3 &  $+$32 09 13 & 0.224 & 8.98$\pm$0.38 & 7.79$\pm$0.37  & 0.43 & $<$0.32$^k$ & 30 (H)& $\surd$\\
RXJ2228.6+2037         &  22 28 34.4 &  $+$20 36 47 & 0.418 & 11.71$\pm$0.20 & 11.71$\pm$0.20  &0.29  & $<$0.95$^v$ & 20 (C)& $\surd$\\
A2537				& 23 08 23.3 & $-$02 11 31 & 0.297&  5.48$\pm$0.14 &  4.54$\pm$0.07 & 0.45 &$<$0.51$^v$ & 39 (C)& $\surd$\\
RXJ0027.6$+$2616 	& 00 27 49.8 & $+$26 16 26 & 0.365 & 3.52$\pm$0.11 &  3.52$\pm$0.11 & 0.24 & $<$0.74$^v$ & 22 (C)\\
Z5699 				& 13 06 00.0 & $+$26 30 58 & 0.306 &	 4.74$\pm$0.08 & 4.74$\pm$0.08 & 0.18 &$<$0.59$^v$ & 26 (C)\\
Z5768				& 13 11 31.5	& $+$22 00 05 & 0.266 & 1.66$\pm$0.06 & 1.66$\pm$0.06 & 0.10 &$<$0.36$^v$ & 27 (C)\\		
S780				& 14 59 29.3 & $-$18 11 13 & 0.236 & 8.68$\pm$0.10  & 6.32$\pm$0.05 & 0.43 &$<$0.38$^v$ & 40 (C)& $\surd$\\
\hline
\sidehead{Radio Halos (EGRHS) }
A2744 				     	&  00 14 18.8 &  $-$30 23 00 & 0.307 &  14.73$\pm$0.24 &  14.73$\pm$0.24 & 0.17 &18.62$\pm$0.94$^a$ & 14 (P)& $\surd$\\    
A0209 					&  01 31 53.0 &  $-$13 36 34 & 0.206 &   7.62$\pm$0.48 & 7.62$\pm$0.48 &0.31 &1.99$\pm$0.21$^a$ & 11 (H)& $\surd$\\
A2163 					&  16 15 46.9 &  $-$06 08 45 & 0.203 &  21.95$\pm$0.33 & 21.95$\pm$0.33 &0.25 & 22.91$\pm$1.16$^a$ & 7 (P)& $\surd$\\
RXCJ2003.5$-$2323 	&  20 03 30.4 &  $-$23 23 05 & 0.317 &   9.17$\pm$0.09 &  9.17$\pm$0.09  &0.09 &10.71$\pm$1.73$^b$ & 50 (C)& $\surd$\\
A520	                           &  04 54 19.0 &  $+$02 56 49 & 0.203  & 7.81$\pm$0.21 &  7.81$\pm$0.21  &0.18 &2.45$\pm$0.18$^a$ & 5 (P)& $\surd$\\
A773                                &  09 17 59.4 &  $+$51 42 23 & 0.217 & 7.30$\pm$0.57  &  7.30$\pm$0.57  & 0.35&1.48$\pm$0.16$^a$ &17 (H)& $\surd$\\
A1758a                            &  13 32 32.1 &  $+$50 30 37 & 0.280 & 8.80$\pm$0.16   & 8.80$\pm$0.16 & 0.18 & 5.75$\pm$0.98$^a$ &16 (P)& $\surd$\\
A2219                             &  16 40 21.1 &  $+$46 41 16 & 0.228 &  14.78$\pm$0.19  &14.78$\pm$0.19 & 0.20 &5.63$\pm$0.80$^a$ & 16 (P)& $\surd$\\
A0521$^U$  					&  04 54 09.1 &  $-$10 14 19 & 0.248 &   8.28$\pm$0.07 & 8.28$\pm$0.07 & 0.08&1.45$\pm$0.13$^i$ & 39 (C)& $\surd$\\
A697$^U$                                &  08 42 53.3 &  $+$36 20 12 & 0.282 & 13.04$\pm$0.61 &  13.04$\pm$0.61 &0.33  &1.51$\pm$0.14$^l$ & 28 (H)& $\surd$\\
A1300$^U$ 					&  11 31 56.3 &  $-$19 55 37 & 0.308 &  11.47$\pm$0.37 & 11.47$\pm$0.37  & 0.18 & 3.8$\pm$1.43$^p$ & 9 (P)& $\surd$\\
\hline
\sidehead{Radio Halos (literature) }
CL0016+16				& 00 18 33.3 &	$+$16 26 36 & 0.541 & 15.54$\pm$0.28  & 15.54$\pm$0.28  & 0.16 &5.01$\pm$0.31$^a$ & 43 (P)& $\surd$\\
A1914					& 14 26 03.0   & $+$37 49 32  & 0.171 &  11.17$\pm$0.13 &10.25$\pm$0.13 & 0.39 & 5.62$\pm$0.43$^a$ & 9 (P)& $\surd$\\
A665					& 08 30 45.2 &	$+$65 52 55 & 0.182  & 8.36$\pm$0.09  &  8.30$\pm$0.07 & 0.22 & 2.51$\pm$0.21$^a$ & 38 (P)& $\surd$\\  
A545					& 05 32 20.2 &	$-$11 31 54 &	0.154 & 6.31$\pm$0.09 & 6.31$\pm$0.09 & 0.23 & 1.41$\pm$0.22$^a$ & 14 (P)& $\surd$\\
Coma					& 12 59 48.7 &	$+$27 58 50 &	0.023  & 3.39$\pm$0.03$^*$ & - & - & 0.76$\pm$0.06$^c$  & -&$\surd$\\
A2256					& 17 03 43.5 &	$+$78 43 03 & 0.058 & 4.44$\pm$0.02 & 4.44$\pm$0.02 &0.22 & 0.85$\pm$0.08$^d$ & 17 (P)& $\surd$\\
Bullet					& 06 58 29.2 &	$-$55 57 10 &	0.296 & 22.54$\pm$0.52   & 22.54$\pm$0.52  & 0.22 & 23.44$\pm$1.51$^e$ & 5 (P)& $\surd$\\
A2255					& 17 12 31.0 & $+$64 05 33 & 0.081 & 3.31$\pm$0.03 &   3.31$\pm$0.03 & 0.14 &0.81$\pm$0.17$^f$ & 15 (P)& $\surd$\\ 
A2319					& 19 20 45.3 &	$+$43 57 43 &	0.056 & 7.96$\pm$0.08 &7.87$\pm$0.08 &0.31 &2.45$\pm$0.19$^g$ & 3 (P)& $\surd$\\
MACS J0717.5+3745	& 07 17 33.8 &	$+$37 45 20 &	0.548 & 24.05$\pm$0.22 & 24.05$\pm$0.22 & 0.17&52.48$\pm$20.56$^h$ & 60 (C)& $\surd$\\
A1995					& 14 52 50.4 &   $+$58 02 48 &	0.319 & 6.03$\pm$0.08 & 6.03$\pm$0.08 &	0.43 &1.66$\pm$0.23$^a$ & 48 (C)& $\surd$\\
MACSJ1149.5+2223$^U$	& 11 49 34.3 &	$+$22 23 42 &	0.544 & 15.50$\pm$0.29  & 15.50$\pm$0.29   & 0.18 & 2.29$\pm$0.95$^m$ & 19 (C)& $\surd$\\
PLCKG171.9-40.7$^U$	& 03 12 57.4 &	$+$08 22 10 & 0.270 & 11.28$\pm$0.02$^{**}$ & -& - & 4.90$\pm$1.35$^n$ &- & $\surd$\\
A754$^U$					& 09 08 50.1 &	$-$09 38 12 & 0.054 &  4.75$\pm$0.033  & 4.75$\pm$0.033  & 0.23 & 0.63$\pm$0.07$^o$ & 8 (P)& $\surd$\\ 
\enddata
\tablecomments{The first part of table contains clusters with radio upper limits belonging to the EGRHS (from Venturi et al. 2008; Kale et al. 2013); the second part clusters with giant RHs belonging to the EGRHS; third part clusters with giant RHs not belonging to the EGRHS; clusters marked with $^U$ are those hosting USSRH ($\alpha>1.5$). 
Columns: (1) Cluster name; (2) e (3) cluster right ascension and declination, respectively, in J2000 coordinates, taken from the X-ray catalogues (see Sect~.2); (4) cluster redshift; (5) 0.1-2.4 keV band cluster X-ray luminosity within $R_{500}$; (6) 0.1-2.4 keV band cluster X-ray luminosity within $R_{500}$ corrected for the contribution of the cool-core; (7) the ratio between the X-ray luminosity within the core and the total luminosity within $R_{500}$, cool-core clusters are indicated with $^{cc}$; (8) k-corrected radio halo power at 1.4 GHz; (9) X-ray exposure in ksec, with P={\it ROSAT} PSPC, H={\it ROSAT} HRI and C= {\it Chandra} ACIS-I; (10) the symbol $\surd$ indicates the clusters present in the 15.5 months {\it Planck} catalogue. Reference for the radio halo powers are: $^v$ Venturi et al. (2008); $^k$ Kale et al. (2013); $^q$ Guglielmino G. (2008 priv. comm.); $^r$ Russell et al. (2011); $^a$ Giacintucci et al. ( in prep.);  $^b$ Giacintucci et al. (2009); $^c$ Kim et al. (1990); $^d$ Clarke \& En\ss lin (2006); $^e$ Liang et al. (2000); $^f$ Govoni et al. (2005); $^g$ Farnsworth et al. (submitted); $^h$ Bonafede et al. (2009); van Weeren et al. (2009); $^i$ Dallacasa et al. (2009); $^l$ Macario et al. (2010); $^m$ Bonafede et al. (2012); $^n$ Giacintucci et al. (2013); $^o$ Macario et al. (2011); $^p$ Venturi et al. (2013).  $^*$ the X-ray luminosity of the Coma cluster is taken from Ohara et al. (2006); $^{**}$ the X-ray luminosity of PLCKG171.9$-$40.7 is taken from Planck Collaboration (2011c).}
\end{deluxetable*}

%
%

\subsection{Identification of cool-core clusters}
\label{sec:coolcore}

In this Section we identify cool-core clusters in our sample to investigate possible biases that can be induced by cool-core clusters on scaling relations and bimodality.

As first measurement, we consider the X-ray surface brightness concentration parameter, defined as the ratio between the X-ray luminosity within the core region ($L_{core}$, within $0.15\,R_{500}$) and $L_{500}$  (\eg Santos et al.  2008; Cassano et al. 2010, with slightly different definitions). In the literature, the concentration parameter has been used for a first identification of cool-core clusters in those cases where a spatially resolved spectroscopic analysis was not possible (\eg in the case of high-redshift clusters; Santos et al. 2008) and to discriminate between merging clusters and more relaxed clusters (\eg Cassano et al. 2010). A large value of this parameter indicates a large probability that the object has a cool core. 

\begin{figure*}
\centering
\includegraphics[width=0.95\textwidth]{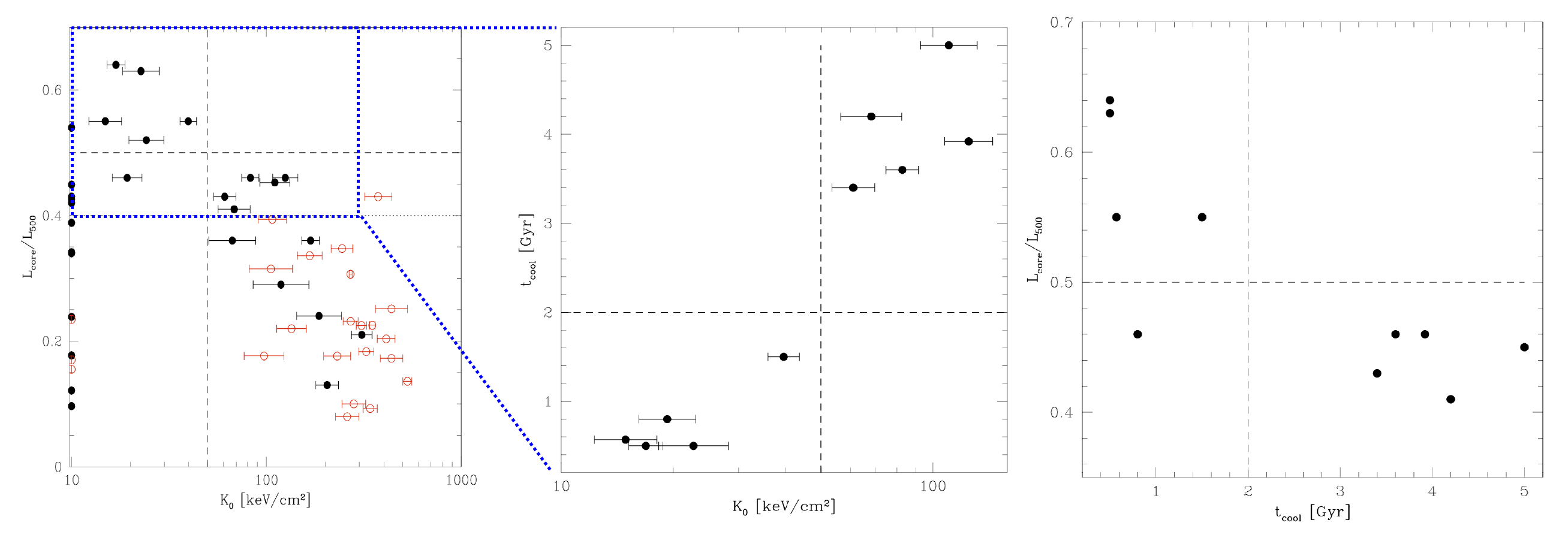}
\caption[]{{\it Left Panel}. $L_{core}/L_{500}$ vs $K_0$ for all clusters in our sample, for those clusters for which we do not find values of $K_0$ in Cavagnolo et al. 2009, we set $K_0=10\,keV/cm^2$ (at the boundary of the plot). Clusters without giant RHs and clusters with giant RHs are reported as black and red points, respectively. {\it Central Panel}. $t_{cool}$ vs $K_0$ for all clusters with $L_{core}/L_{500}>0.4$ and $K_0<130\,keV/cm^2$ (blue dashed region in the {\it Left Panel}); {\it Right Panel}. $L_{core}/L_{500}$ vs $t_{cool}$ for all clusters with $L_{core}/L_{500}>0.4$ and $K_0<130\,keV/cm^2$.}
\label{Fig.conc}
\end{figure*}

The derived values of the ratio $L_{core}/L_{500}$ are reported in Table~\ref{Tab.ULRH}. Here we use the concentration parameter in combination with the central entropy ($K_0$ in $\mathrm{keV/cm^2}$) and the central cooling time ($t_{cool}$) to identify clusters with a cool-core. Values of $K_0<50\, \mathrm{keV/cm^2}$ (dashed vertical line in Fig.~\ref{Fig.conc}) are used to identify cool-core clusters (\eg Cavagnolo et al 2009; Rossetti et al. 2011).

We inspected the sample of Cavagnolo et al. (2009) to find information about the central entropy for our clusters, and we report in Fig.~\ref{Fig.conc} ({\it left panel}) the distribution of clusters in the $L_{core}/L_{500}$ vs $K_0$ diagram; 13 of our clusters are not available in the Cavagnolo et al. sample and in Fig.~\ref{Fig.conc} ({\it left panel}) they are reported with a value of $K_0=10\,\mathrm{keV/cm^2}$. We find that:

\begin{itemize}
\item[-] clusters with giant RHs (open red dots) have $L_{core}/L_{500}<0.4$\footnote{the only exception is A1995 with $L_{core}/L_{500}=0.43$ (see also discussion in Sect.6.2).} and $K_0>90\, \mathrm{keV/cm^2}$.

\item[-] 5 clusters have $K_0<50\, \mathrm{keV/cm^2}$ and $L_{core}/L_{500}>0.5$: RXJ1532.9+3021, RXCJ1115.8 $+$ 0129, Z2089, RXJ0439.0$+$0520, Z2701, and one, A2667, has $L_{core}/L_{500}=0.46$;

\item[-] 5 clusters have $50 \, \mathrm{keV/cm^2}<K_0<130\, \mathrm{keV/cm^2}$ and $L_{core}/L_{500}>0.4$: A611, A2261, A3088, A1423 and A2537. 

\end{itemize}

\noindent As expected, clusters with giant RHs, can be easily identified with {\it merging} clusters. To better understand whether  the 11 clusters with $L_{core}/L_{500}>0.4$ and $K_0<130\, \mathrm{keV/cm^2}$ are cool-core or non cool-core clusters, we searched for information in the literature about their central cooling time ($t_{cool}$) (Fig.~\ref{Fig.conc}, {\it central panel})\footnote{we do not found information about $t_{cool}$ in the literature for Z2089.}.  Clusters with $K_0<50 \, \mathrm{keV/cm^2}$ have $t_{cool}< 2$ Gyr, while the others have $t_{cool}>3$ Gyr. Fig.~\ref{Fig.conc}, {\it right panel}, also shows that clusters with $K_0<50 \, \mathrm{keV/cm^2}$ and $t_{cool}< 2$ Gyr all have $L_{core}/L_{500}>0.5$ (with the exception of A2667).
Therefore, based on the combination of the three indicators, we identify cool-core clusters in our sample as those with $L_{core}/L_{500}>0.5$, \ie clusters which emit more than 50\% of their $L_{500}$ within their cores \footnote{we also consider A2667 as a cool-core cluster, since it has an estimated central entropy of $K_0\approx 19$ keV/cm$^2$ and a central cooling time $t_{cool}\approx 1$ Gyr and it is classified as a cool core cluster by Zhang et  al. (2007).}. We can thus conclude that in our sample there are 7 cool-core clusters\footnote{The cluster Z348 has no information about $K_0$ in the literature, but since it has $L_{core}/L_{500}=0.54$ we can identify it as a cool-core cluster.} (these are marked with a $^{cc}$ symbol in Col.~7 of Table~\ref{Tab.ULRH}).

\subsection{Cluster Sunyaev-Zel'dovich measurements}
\label{Sect.SZ}

Observations of clusters through their SZ-effect offer a valid alternative to X-rays for the measure of the cluster mass, since the magnitude of the SZ effect is proportional to the integral along the line of sight of the cluster pressure, and hence is proportional to the cluster mass. The total SZ signal can be defined as

\begin{eqnarray}
\label{Eq.SZ}
Y_{\Delta_c}=D_A^2 Y_{SZ}=(\frac{\sigma_T}{m_ec^2})\int_{R\leq R_{\Delta_c}} PdV \,\propto \nonumber\\
M_{gas} T_e=f_{gas} M_{tot} T_e
\end{eqnarray}

\noindent where $D_A$ is the angular diameter distance to the system, $\sigma_T$ the Thomson cross-section, $c$ the light speed, $m_e$ the electron rest mass, $P=n_ekT_e$ the electron pressure, $f_{gas}$ is the gass mass fraction and $M_{tot}$ the total cluster mass. The integral in Eq.~\ref{Eq.SZ} is performed over a sphere of radius $R_{\Delta_c}$, which is the radius corresponding to a density contrast $\Delta_c\,\rho_c(z)$. 
When the integration is performed over a sphere of radius $R_{500}$ the SZ signal is denoted with $Y_{500}$, that in the following of the paper will have the unit dimension of Mpc$^2$.

For all clusters in Table~\ref{Tab.ULRH} we search for information about the SZ signal in the recent all-sky {\it Planck} SZ cluster catalogue (PSZ), which contains all validated clusters from the first 15.5 months of {\it Planck} satellite observations (Planck Collaboration 2013b). Considering only clusters belonging to the EGRHS sub-sample we find that 11/11 RH clusters and 19 out of 29 clusters with upper limits are contained in the PSZ catalogue. Among the 10 clusters not present in the PSZ catalogue 5 are cool-core clusters, therefore only 2 out of 7 cool-core clusters of our sample are detected by {\it Planck}. The remaining 14 RH clusters from the literature are also contained in PSZ catalogue. 

We obtain a sub-sample of 44 clusters (25 halos and 19 upper limits) for which {\it Planck} measurements of $Y_{500}$ are available (see Table~\ref{Tab.YSZ_YX}).

%
%
\begin{deluxetable}{llcc}
\tabletypesize{\scriptsize} 
\tablecolumns{4} 
\tablewidth{0pc} 
\tablecaption{Observed cluster SZ properties\label{Tab.YSZ_YX}}
\tablehead{
\colhead{cluster name}  &  \colhead{index} & \colhead{$log(Y_{500})$} &  \colhead{$log(M_{500})$} \\
                  \colhead{}                       &  \colhead{}         &      \colhead{$\mathrm{Mpc^2}$} &  \colhead{$M_{\odot}$}}
\startdata
\sidehead{Upper limits (EGRHS) }
 A2697	& 315  & -4.150$\pm$0.077 &  14.78$\pm$0.04\\
 A141 				   & 599  & -4.379$\pm$0.120 &  14.65$\pm$0.07\\
 A3088				   & 744  & -4.062$\pm$0.065 &  14.83$\pm$0.04\\
								RXCJ1115.8$+$0129  & 881  &-4.087$\pm$0.087  &  14.80$\pm$0.05\\
 								 A2631				   & 297  &-4.029$\pm$0.067  &  14.84$\pm$0.04\\
								 A2645				   & 254  &-4.288$\pm$0.099	 &  14.70$\pm$0.06\\
								 A2667				   & 94    &-4.054$\pm$0.055  &  14.83$\pm$0.03\\
								RXJ0142.0$+$2131 	   & 500  &-4.134$\pm$0.102  &  14.78$\pm$0.06\\
								A267					   & 541  &-4.301$\pm$0.108  &  14.69$\pm$0.06\\
								RXJ0439.0$+$0715	   & 640  &-4.181$\pm$0.096	 &  14.76$\pm$0.05\\
								A611					   & 623  &-4.162$\pm$0.081   &  14.77$\pm$0.05\\
								 A781					   & 654  &-4.097$\pm$0.072   &  14.80$\pm$0.04\\      
								A1423				   & 610  &-4.143$\pm$0.064   &  14.78$\pm$0.04\\
								A1576				   & 460  &-4.143$\pm$0.063	  &  14.78 $\pm$0.04\\           
 								A2146				   & 359  &-4.495$\pm$0.080	  & 14.58$\pm$0.05\\
								A2261				   & 174  &-3.991$\pm$0.048	  & 14.87$\pm$0.03\\
								RXJ2228.6$+$2037 	   &	275  &-3.917$\pm$0.072    & 14.89$\pm$0.04\\
								A2537				   &	247  &-4.120$\pm$0.080    & 14.79$\pm$0.04\\
								S780					   &1185 &-3.957$\pm$0.062	  &  14.89$\pm$0.03\\
								S780					   &1185 &-3.957$\pm$0.062	  &  14.89$\pm$0.03\\
\sidehead{Radio Halos (EGRHS) }
A2744  	                & 26   & -3.778$\pm$0.041 & 14.98$\pm$ 0.02\\
								   A0209   			    & 558 & -3.916$\pm$0.041 & 14.91$\pm$ 0.02\\
								   A2163				    & 19   & -3.374$\pm$0.019  &15.22$\pm$ 0.01 \\
								  RXCJ2003.5$-$2323 & 46   & -3.967$\pm$0.068 & 14.87$\pm$0.04 \\
								  A520				    & 655 &-4.030$\pm$0.062  & 14.85$\pm$0.04\\
 								  A773				    & 578 &-4.026$\pm$0.049  & 14.85$\pm$0.03\\
								 A1758a				    & 389 &-3.922$\pm$0.044  & 14.90$\pm$0.03\\
								 A2219				    & 242 &-3.681$\pm$0.026  & 15.04$\pm$0.01\\
								 A521$^U$					   &	688  & -4.040$\pm$0.070	  & 14.83$\pm$0.04\\
								 A697	$^U$				   & 628 & -3.640$\pm$0.032 	  & 15.06$\pm$0.02\\
								A1300$^U$				   & 960  & -3.839$\pm$0.053  & 14.95$\pm$0.03\\
\sidehead{Radio Halos (literature) }
CL0016$+$1609         & 408  &-3.813$\pm$0.077  & 14.94$\pm$0.04\\
								  A1914				   &  224 &-4.045$\pm$0.039   & 14.84$\pm$0.02\\
								A665					   &  533 &-3.914$\pm$0.037   & 14.92$\pm$0.02\\
								 A545					   &  707 &-4.397$\pm$0.112   & 14.64$\pm$0.06\\
								 Coma				   & 187  &-4.281$\pm$0.030   & 14.72$\pm$0.02\\
								 A2256				   & 407 & -4.135$\pm$0.022   & 14.80$\pm$0.01\\
								A2255				   & 325 & -4.288$\pm$0.028	  & 14.71$\pm$0.02\\  
								A2319				   & 252 & -3.900$\pm$0.020   & 14.93$\pm$0.01\\
								MCSJ0717.5$+$3745 & 608 & -3.612$\pm$0.049   & 15.05$\pm$0.03\\
								 Bullet				   & 920 & -3.577$\pm$0.025	  & 15.09$\pm$0.02\\
								 A1995				   & 337 & -4.257$\pm$0.075	  & 14.71$\pm$0.04\\
								MCSJ1149.5$+$2223$^U$  & 765 & -3.824$\pm$0.072   & 14.93$\pm$0.04\\
								PLCK G171.9$-$40.7$^U$   &  591 & -3.666$\pm$0.039  & 15.05$\pm$0.02\\
								 A754$^U$ 					   &  801 &-4.095$\pm$0.023   & 14.82$\pm$0.01 \\
\enddata
\tablecomments{Columns: (1) radio properties; (2) cluster name; (3) index indicating the position in the {\it Planck} validation catalogue; (4) logarithmic value of $Y_{500}$ in Mpc$^2$, with 68\% errors;  (5) logarithmic value of $M_{500}$ in solar masses, with 68\% errors. Clusters marked with $^U$ are those hosting USSRHs.
All the $M_{500}$ and $Y_{500}$ values refer to Planck Collaboration (2013b) (from the website: http://szcluster-db.ias.u-psud.fr);  with the exception of the cluster PLCK G171.9$-$40.7 whose values are taken from Planck Coll. (2011c).}							
\end{deluxetable}

%
%

For the same clusters we also find information in the PSZ catalogue about the values of $M_{500}$. These are obtained from $Y_{500}$ as described in Planck Collaboration (2013b; Sect. 7.2.2) and are reported in Table~\ref{Tab.YSZ_YX}.

\subsection{Cluster dynamical status}
\label{Sect.dyn}

For clusters belonging to the EGRHS with information about $Y_{500}$ and $M_{500}$ (see Table~\ref{Tab.YSZ_YX}) we make use of {\it Chandra} archival data to determine the cluster dynamical status\footnote{With the only exception of Abell 2697 for which {\it Chandra} data are not available.}.
We produce X-ray images in a standard manner using CIAO 4.3 (with calibration files from CALDB 4.4.1) in the 0.5-2 keV band. We adopted an algorithm for an automatic detection of the point sources, that are then removed from the images.
Following Cassano et al. (2010), we study the cluster substructure on the RH scale analyzing the surface brightness inside an aperture radius of 500 kpc, since we are interested in the cluster dynamical properties on the scales where the energy is most likely dissipated. We use two methods: the emission centroid shift (e.g., Mohr et al. 1993; Poole et al. 2006, OÕHara et al. 2006; Ventimiglia et al. 2008, Maughan et al. 2008, B\"ohringer et al. 2010) and the surface brightness concentration parameter (e.g., Santos et al 2008).

The centroid shift, $w$, is computed in a series of circular apertures centered on the cluster X-ray peak and is defined as the standard deviation of the projected separation between the peak and the centroid in unit of $R_{ap}$, as (Poole et al. 2006; Maughan et al. 2008):

\begin{equation}
w=\Big[\frac{1}{N-1}\sum (\Delta_i-\langle \Delta \rangle)^2\Big]^{1/2}\times \frac{1}{R_{ap}},
\label{Eq:w}
\end{equation}

\noindent where $\Delta_i$ is the distance between the X-ray peak and the centroid of the {\it i}th aperture.
\vskip 0.3cm


Following Santos et al. (2008), the concentration parameter, $c$, is defined as the ratio of the peak over the ambient surface brightness, $S$, as:

\begin{equation}
c=\frac{S(r<100\,\mathrm{kpc})}{S(<500\,\mathrm{kpc})}
\label{Eq:c}
\end{equation}

\noindent 
We use the concentration parameter to differentiate galaxy clusters with a compact core (\ie core not disrupted from a recent merger event) from clusters with a spread distribution of gas in the core (\ie core disturbed from a recent merger episode). 

Cassano et al. (2010) showed that considering the median value of each parameter, $w=0.012$ and $c = 0.2$, it was possible to separate the sample between RH merging clusters ($w>0.012$ and $c<0.2$) and more relaxed clusters without RHs ($w<0.012$ and $c>0.2$). We will use these values as reference for our sample.

\subsection{Fitting procedure}
\label{Sect.method}

Here we describe the procedure used in next Sections to investigate the presence of scaling relations between independent measurements, \ie the RH power and the cluster thermal quantities ($L_{500}$,  $L_{500,cor}$, $Y_{500}$ and $M_{500}$). For each set of observables we fit a power-law relation using linear regression in the log-log space by adopting the BCES-bisector and the BCES-orthogonal regression algorithms (Akritas \& Bershady 1996) which treat the variables symmetrically and take into account measurement errors in both variables and intrinsic scatter in the data. Performing Monte Carlo simulations to test the performances of different regression methods, Isobe et al. (1990) recommended the use of BCES-bisector in the case one would like to treat the variables symmetrically. Consequently, we will consider the BCES-bisector as reference method. 

Since we also have upper limits on $P_{1.4}$, in those cases where upper limits and detections are not clearly separated we also use a regression analysis based on the parametric EM ({\it Expectation-Maximization}) algorithm that is implemented in the {\it ASURV} package (Isobe, Feigelson \& Nelson 1986) and deals with ``censored data'', upper limits.

Assuming a linear relation of the form $Y = aX+b$, and a sample of $N$ data points ($Y_i, X_i$) with errors $\sigma_{Y{i}}$ and $\sigma_{X{i}}$, we estimate the raw scatter using the error weighted orthogonal distances to the regression line (\eg Pratt et al. 2009; Biffi et al. 2013): 

\begin{equation}
\sigma^2_{raw}=\frac{1}{N-2}\sum_{i=1}^{N}w_i(Y_i-aX_i-b)^2
\end{equation}

\noindent where 

\begin{equation}
w_{i}=\frac{1/\sigma_i^2}{(1/N)\sum_{i=1}^{N}1/\sigma_i^2} \ \ \ \ \ \ \ {\rm and}\ \ \ \ \ \ \sigma_i^2 = \sigma^2_{Y_i} + a^2 \sigma^2_{X_i}.
\end{equation}


Since we are dealing with a limited sample, the regression line obtained for our data is a {\it sample regression line} that can deviate from the (unknown) {\it true regression line}. To evaluate the variation of our best-fit relation about the true regression line, we estimate the 95\% confidence interval for the mean value of $<Y>$ at a given $X$,  \ie the area that has a 95\% chance of containing the true regression line. 
For a given value of the $X$ variable the 95\% confidence region around the mean $<Y>$ (which is given by the best-fit relation: $<Y>=aX+b$) is $<Y>\pm\Delta Y$, where:

\begin{equation}
\Delta Y=\pm 1.96 \sqrt{\sum_{i=1}^{N} \frac{(Y_i-Y_m)^2}{N-2}}\sqrt{\Bigg(\frac{1}{N}+\frac{(X-X_m)^2}{\sum_{i=1}^{N} (X_i-X_m)^2}\Bigg)}
\end{equation}

\noindent where for each observed $X_i$, $Y_m=aX_i+b$, and $X_m=\sum_{i=1}^{N} X_i/N$.

 \section{Expected scaling relations}
\label{Sect.models}

Scaling relations between the synchrotron radio power of halos and the cluster thermal properties (mass, X-ray luminosity, temperature) are expected in theoretical models for the formation of giant RHs. 
In this Section we briefly summarize the basic theoretical expectations for the scalings.  

\subsection{Secondary models}
In the simplest scenario for the formation of giant RHs in clusters, the electrons responsible for the synchrotron emission are secondary products of the hadronic interaction between thermal and cosmic ray protons. In this model, following the formalism by Kushnir et al. (2009), the scaling between the synchrotron radio power and the cluster [0.1-2.4] keV X-ray luminosity is expected to be $\nu\,P_{\nu}^{syn}\propto L_X^{\frac{\alpha_L+0.5}{\alpha_L-0.6}}$, where $\alpha_L$ is the slope of the L-T relation. For $\alpha_L\simeq 2.5-3$ (\eg Markevitch 1998; Arnaud \& Evrard 1999; Reiprich \& Bohringer 2002; Pratt et al. 2009), one obtains:

\begin{equation}
\nu\,P_{\nu}^{syn}\propto L_X^{1.58-1.46}
\end{equation}
 
\noindent This is valid under the assumption that the relevant radiation losses for the secondary electrons are synchrotron losses, \ie assuming that the average magnetic field strength in the halo volume is $B>B_{CMB} \simeq 3.2(1+z)^2\mu$G. Lower magnetic field values are disfavored by the combination of {\it Planck} and {\it Fermi} data with radio observations (\eg Jeltema \& Profumo 2011; Planck Collaboration 2013a; Brunetti et al. 2012).

\noindent Since $Y_{500}$ is found to scale as $Y_{500}\propto L_X^{1.02\pm0.07}$ (\eg Planck Collaboration et al. 2011b), this model predicts:

\begin{equation}
\nu\,P_{\nu}^{syn}\propto (Y_{500})^{1.55-1.43}
\end{equation}

\subsection{Turbulent re-acceleration models}

In the case of the turbulent re-acceleration scenario the derivation of scaling relations is less straightforward, due to our poor knowledge of the details of the microphysics of the ICM.
A simple approach to derive scaling relations in this model is presented in Cassano et al. (2007). Under quasi-stationary conditions, the energy flux of the turbulence which goes into relativistic electrons is reradiated via synchrotron and IC mechanisms. The injection rate of the turbulence generated during a merger in the RH volume can be estimated as $\dot{\varepsilon_t} \propto \rho_H\times v^2/\tau_{cros}$, where  $\rho_H$ is the ICM mean density in the RH volume,  $v$ is the cluster-cluster impact velocity and $\tau_{cros}$ is the cluster crossing time. As in the case of secondary models, it is assumed that the ratio between the energy densities in relativistic particles and thermal plasma does not change in any systematic way with cluster mass (or temperature) among RH clusters. Under this hypothesis the synchrotron radio power is $\nu\,P_{\nu}^{syn}\propto (M_H\sigma_H^3)/{\mathcal{F}(z,M_H,b_H)}$, where $\mathcal{F}(z,M_H,b_H) =[1+(3.2 (1+z)^2/B_H)^2]$, and $M_H$,  $\sigma_H$ and $B_H$ are the total cluster mass, the cluster velocity dispersion and the average magnetic field strength within the RH size ($R_H$), respectively (Cassano et al. 2007). The expression $\mathcal{F}$ is constant in the asymptotic limit $B_H^2>>B_{cmb}^2$, or  when the magnetic field in the RH region is independent of the cluster mass. In this case, $\nu\,P_{\nu}^{syn}\propto M_H^{1.8}$. Assuming the scalings $M_H\propto R_H^{2.17}$ (Cassano et al. 2007) and $R_H\propto R_{500}^{3.1}$ (Basu 2012), one has:

\begin{equation}
\nu\,P_{\nu}^{syn}\propto M_{500}^{4.0}
\end{equation}

\noindent and considering the scaling $M_{500}\propto Y_{500}^{1/1.74}$ one has:

\begin{equation}
\nu\,P_{\nu}^{syn}\propto Y_{500}^{2.3}
\end{equation}

\noindent that is steeper than that predicted by ``secondary models''. Re-acceleration modes also allow the case $B_H^2 << B_{cmb}^2$, without tension with $\gamma$-ray upper limits (\eg Brunetti et al. 2012), and in this case one has $\mathcal{F}^{-1} \propto M_H^{2 b_H}$, 
that implies a correlation even steeper than that obtained in the previous case.

Besides the details of the slopes of the thermal-non-thermal scaling relations expected from a different origin of the emitting electrons, an important difference between the two scenarios is the expected dispersion of the correlations. Re-acceleration models predict a variety of spectral shapes of RHs, including very steep spectra (\eg Cassano et al. 2006; Brunetti et al. 2008), which imply a substantial dispersion in the correlations (Kushnir et al. 2009; Brunetti et al. 2009) and an increase of the scatter at low observing frequency (Cassano 2010).

\section{Radio--X-ray luminosity correlation and the bimodality}
\label{sec:radioX}

It is well known that the radio luminosity of halos at 1.4 GHz scales with the X-ray luminosity of the hosting clusters (\eg Liang et al. 2000; Feretti 2002, 2003; En\ss lin \& R\"ottgering 2002; Cassano et al. 2006; Brunetti et al. 2009; Giovannini et al. 2009). This correlation has been used to claim that a correlation should exist also between the radio power and the virial mass of the host cluster (\eg Cassano et al. 2006). 
Deep upper limits to the radio flux density of clusters with no RH emission at 610 MHz, which were a factor of $\sim3\div20$ below the correlation, were obtained from the GRHS and its extension\footnote{Previous attempts to compare upper limits and the correlation can be found in Dolag (2006).} allowing to validate the correlation itself and to discover the radio bimodality (\eg Brunetti et al. 2007). 
						    
In previous papers, the distribution of galaxy clusters in the radio-X-ray luminosity diagram, and the scaling relation between the two quantities, were based on non-homogeneous radio and X-ray measurements. In particular, the radio luminosities of halos were collected from the literature and X-ray luminosities were taken from RASS-based cluster catalogues. Here we recomputed the radio flux densities of well known RHs by reanalyzing observations from the archives (as outlined in Sect.\ref{Sect.radio}). For all clusters we computed the 0.1-2.4 keV X-ray luminosities within $R_{500}$ from pointed {\it ROSAT} and {\it Chandra} observations (see Sect.\ref{Sect.Lx}).

\begin{figure*}
\begin{center}
\includegraphics[width=0.42\textwidth]{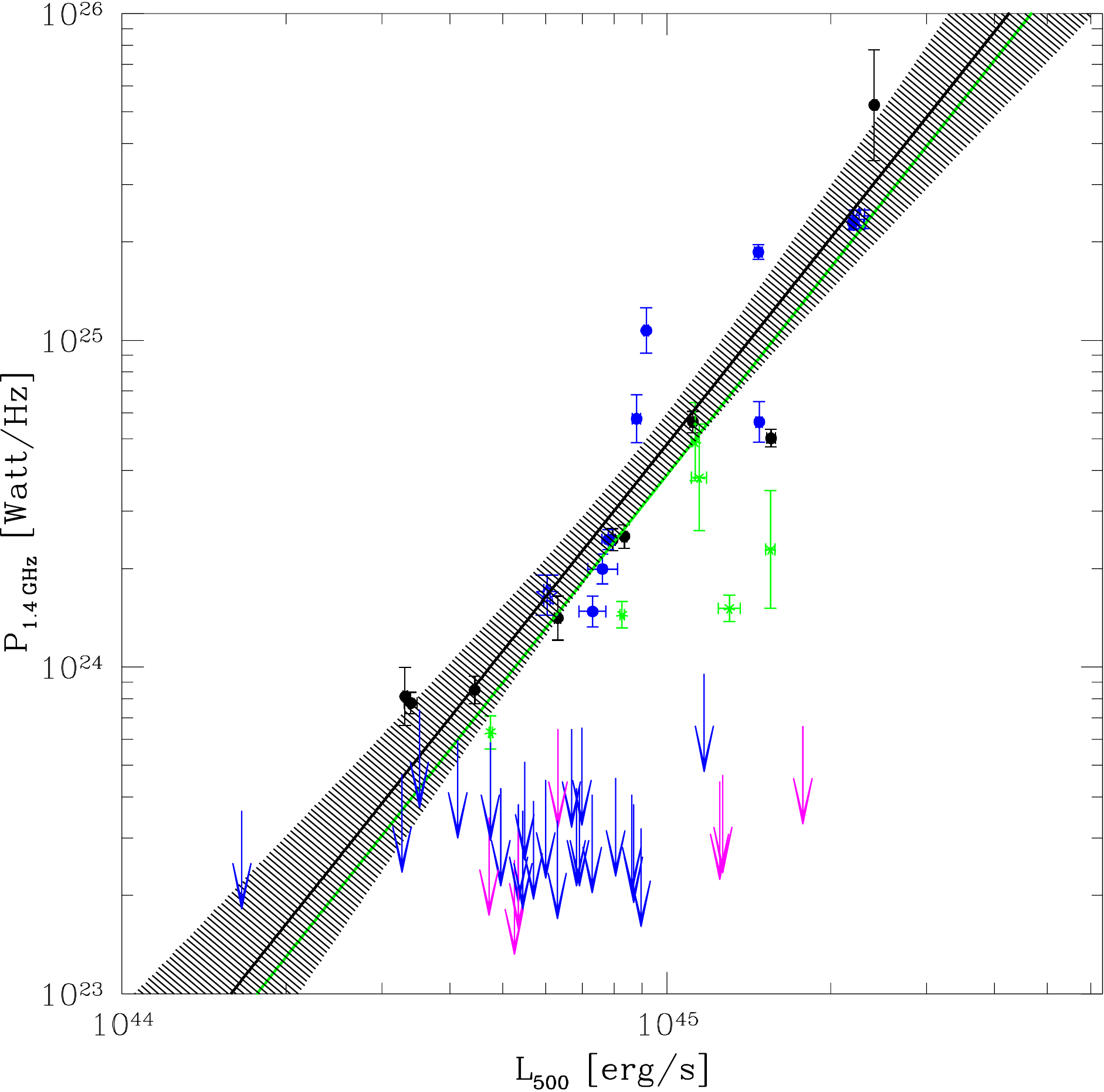}
\includegraphics[width=0.42\textwidth]{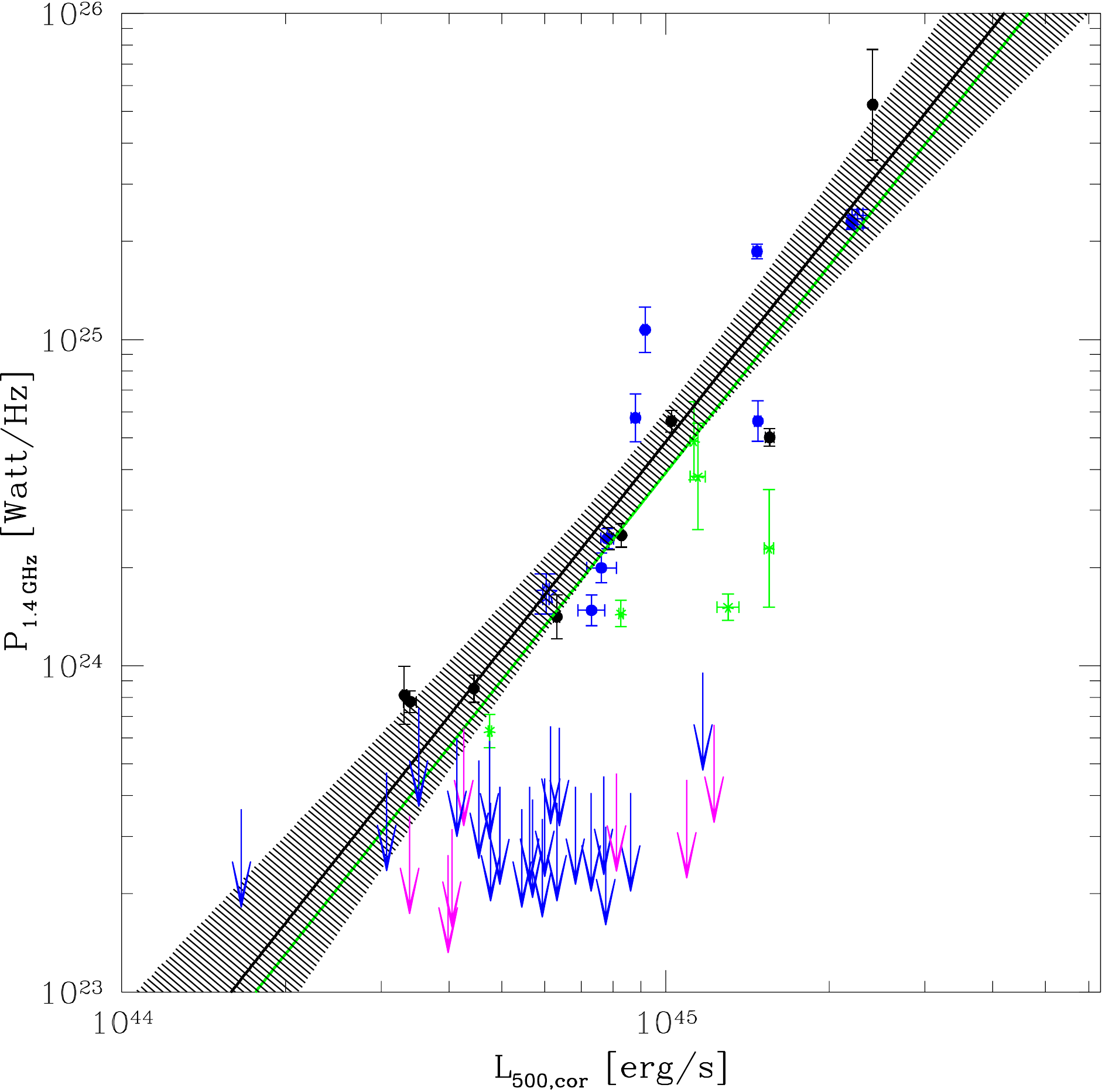}
\caption[]{{\it Left Panel}. Distribution of clusters in the $P_{1.4} - L_{500}$ plane. {\it Right Panel}. Distribution of clusters in $P_{1.4} - L_{500,cor}$ plane. In both panels different symbols indicate: halos belonging to the EGRHS (blue filled dots); halos from the literature (black open dots); halos with very steep spectra (USSRH, green asterisks); A1995 and Bullet cluster (blue stars); cool core clusters belonging to the EGRHS (magenta arrows). Best-fit relations to giant RHs only (black lines) and to all RHs (including USSRH, green dashed lines) are reported. The 95\% confidence regions of the best-fit relations obtained for giant RHs only are also reported (shadowed regions).}
\label{Fig.Lr_Lx}
\end{center}
\end{figure*}

In Fig.~\ref{Fig.Lr_Lx}, ({\it left panel}) we show the distribution of clusters in the $P_{1.4}-L_{500}$ diagram. We report with different colors clusters belonging to the EGRHS (blue points and blue and magenta arrows) and halos from the literature (black points). This is necessary, since the comparison between RH powers and upper limits 
makes sense only for those clusters observed within the same redshift range, and this is possible only for clusters belonging to the EGRHS. Halos from the literature follow the same distribution of halos from the EGRHS, and thus we use them to draw the correlation.
RH clusters appear to follow a well-defined correlation

%
%
\begin{deluxetable*}{lccccccc}
\tabletypesize{\scriptsize} 
\tablecolumns{8} 
\tablewidth{0pt}
\tablecaption{Best-fit parameters of scaling relations\label{Tab.bestfit}}
\tablehead{
\colhead{method}  & \colhead{B}  &  \colhead{err(B)} & \colhead{A}   &  \colhead{err(A)} & \colhead{$\sigma_{raw}$} & \colhead{$r_s$} & \colhead{$P$}}
\startdata
\cutinhead{$P_{1.4}-L_{500}$}
RH+USS                 &         &             &             &            &           &          &                                  \\
BCES Bisector       & 2.11 &   0.20 &  0.088  & 0.056 & 0.23 & 0.83 & $2.32\times10^{-7}$\\
bootstrap             & 2.11  &  0.21  &  0.083  & 0.058 & \\
								     BCES Orthogonal &  2.35 &  0.25 &   0.094 &  0.058 \\
								        bootstrap    	     &  2.37  & 0.31  & 0.089     &  0.062\\					
\\								     
								     
								     RH only & & & & \\
								     BCES Bisector   & 2.10 &   0.17 &  0.181  & 0.048 & 0.20 & 0.95 & $1.03\times10^{-7}$\\
								     bootstrap       & 2.11  &  0.19  &  0.176  & 0.049 & \\
								     BCES Orthogonal  & 2.20 &   0.18 &   0.185 &  0.049\\
								    bootstrap    			& 2.21 &   0.23 &   0.180 &  0.049\\
								    
\cutinhead{$P_{1.4}-L_{500,cor}$} 
RH+USS & & & & &\\
							BCES Bisector &  2.11 & 0.20 &  0.091 &   0.056 & 0.23 & 0.83 & $2.32\times10^{-7}$\\
								   bootstrap  & 2.12 &  0.22 &   0.085  & 0.060 & \\
								 BCES Orthogonal &  2.35  & 0.25  & 0.098 &   0.058\\
								  bootstrap		  &  2.38 &  0.31 &  0.094 &  0.065\\
								\\
								 RH only & & & & \\
								 BCES-Bisector           & 2.11 & 0.16 & 0.186 & 0.048 & 0.20 & 0.95 & $1.03\times10^{-9}$\\
								 bootstrap	                 & 2.11 & 0.18 & 0.184 & 0.050 &\\
								 BCES Orthogonal  & 2.20 &  0.18 &  0.190 &  0.049\\
								 bootstrap 		   &	2.22 &  0.22 &  0.187 &   0.052\\
\cutinhead{$P_{1.4}-M_{500}$}
RH+USS & & & & & \\
								     BCES Bisector &  3.70 &   0.56 &   0.009 &  0.074 & 0.37 & 0.73 & $3.98\times10^{-5}$\\
 								     bootstrap       & 3.73 &   0.64 &   0.011 &   0.079 \\
								     BCES Orthogonal &   5.05 &  0.99  & 0.002  & 0.094\\ 
								    	 bootstrap       		& 5.27 &   1.33 & -0.002 &  0.107\\							    
								    \\
								     RH only & & & & \\
								     BCES-Bisector              & 3.77 & 0.57 & 0.125 & 0.076 & 0.35 & 0.81 & $2.50\times10^{-5}$ \\
								    bootstrap	                    & 3.84 & 0.66 & 0.126 & 0.079\\
								    BCES Orthogonal   &  4.51 &  0.78 &   0.129   & 0.087\\
								   	bootstrap			&   4.62 &   0.90 & 0.131   & 0.092\\	
\cutinhead{$P_{1.4}-Y_{500}$}
 RH+USS & & & & &\\ 
								 BCES Bisector &  2.02 & 0.28 &  -0.131 &   0.070 & 0.35 & 0.74 & $2.66\times10^{-5}$\\
								   bootstrap  & 2.03 &  0.30 &   -0.133  & 0.069 &\\
								 BCES Orthogonal  & 2.48 &  0.43  & -0.167 &  0.089\\
  								    bootstrap   		   &	2.55 &  0.51 & -0.177 &   0.100\\
\\
								 RH only & & & & &\\
								 BCES-Bisector           & 2.05 & 0.28 & -0.014 & 0.068 &0.32 & 0.83 & $1.26\times10^{-5}$\\
								 bootstrap	                    & 2.07 & 0.30 & -0.016 & 0.072 \\
								 BCES Orthogonal     & 2.28 & 0.35 & -0.027 & 0.073 \\
								 bootstrap 		      & 2.30 & 0.38 & -0.030 & 0.079\\	
\\
								 RH+UL  & & & & &\\
								 EM algorithm &  2.77 & 0.54 &  -0.55 &   0.13 & \\
\\
								 RH only & & & & &\\
								   EM algorithm   & 1.70 & 0.26 & 0.006 & 0.068 & \\								    
\enddata
\tablecomments{The last two columns gives the SpearmanÕs rank correlation coefficient, $r_s$, and the related probability of no correlation.}
\end{deluxetable*}
%
%

\noindent between the halo radio power and $L_{500}$. Being steeper than other halos, ultra-steep spectrum RH (green asterisks) are in general under-luminous with respect to this correlation. We remind that the position of USSRH in the $P_{1.4}-L_{500}$ diagram cannot be compared with that of the upper limits as the latter were scaled at 1.4 GHz using $\alpha=1.3$.
We find a bimodal distribution of clusters with the presence of two distinct populations, that of radio-halo clusters and that of radio-quiet clusters. For values of $L_{500}\gtsim5\times10^{44}$erg/s, clusters with upper limits to the radio power (blue and magenta arrows) are all located below the 95\% confidence region of the correlation. 

As the EGRHS is based on X-ray-selected clusters, one may suspect that the bimodality could be caused by the presence of cool-core clusters, which are brighter in X-ray and do not host giant radio-halos. With the idea to test the bimodality against the presence of cool-core clusters in the EGRHS, we derive the distribution of clusters in the $P_{1.4}-L_{500,cor}$ diagram (Fig.~\ref{Fig.Lr_Lx}, {\it right panel}). We highlight the position of cool-core clusters (identified as outlined in Sect.\ref{sec:coolcore}, magenta arrows in Figs.~\ref{Fig.Lr_Lx}). As expected, the X-ray luminosity of cool-core clusters is significantly reduced going from $L_{500}$ to $L_{500,cor}$.

However, the bimodal behavior in the halo radio power remains also in the $P_{1.4}-L_{500,cor}$  diagram. Also in this case, if we restrict to clusters with $L_{500,cor}\gtsim5\times 10^{44}$ erg/sec, upper limits are all below the 95\% confidence region of the correlation. We may thus conclude that the observed radio bimodality is not driven by the presence of cool-core clusters without diffuse radio emission in the EGRHS. 
We fit the observed $P_{1.4}-L_{500}$  and $P_{1.4}-L_{500,cor}$ relation with a power-law of the generic form:

\begin{equation}
\log\Big(\frac{P_{1.4}}{10^{24.5}\mathrm{Watt/Hz}}\Big)=B\,\log\Big(\frac{L_X}{10^{45}\mathrm{erg/s}}\Big)+A
\end{equation}

\noindent where $L_X$ is $L_{500}$ or $L_{500,cor}$. The fit was performed using linear regression in the log-log space by adopting both the BCES-bisector and BCES-orthogonal methods (as discussed in Sect.~\ref{Sect.method}). The results of the fit, together with that from 1000 bootstrap resamples, are reported in Table~\ref{Tab.bestfit}. The slope of the correlation is $\sim 2.1\pm0.2$ and $\sim 2.2\pm0.2$, in the BCES-bisector and  BCES-orthogonal cases, respectively, consistent with that found in previous studies (\eg Brunetti et al. 2009). 
The best-fit relation has a lower normalization and a larger $\sigma_{raw}$ when USSRH are included in the fit (see Table~\ref{Tab.bestfit}).

\begin{figure*}
\begin{center}
\includegraphics[width=0.45\textwidth]{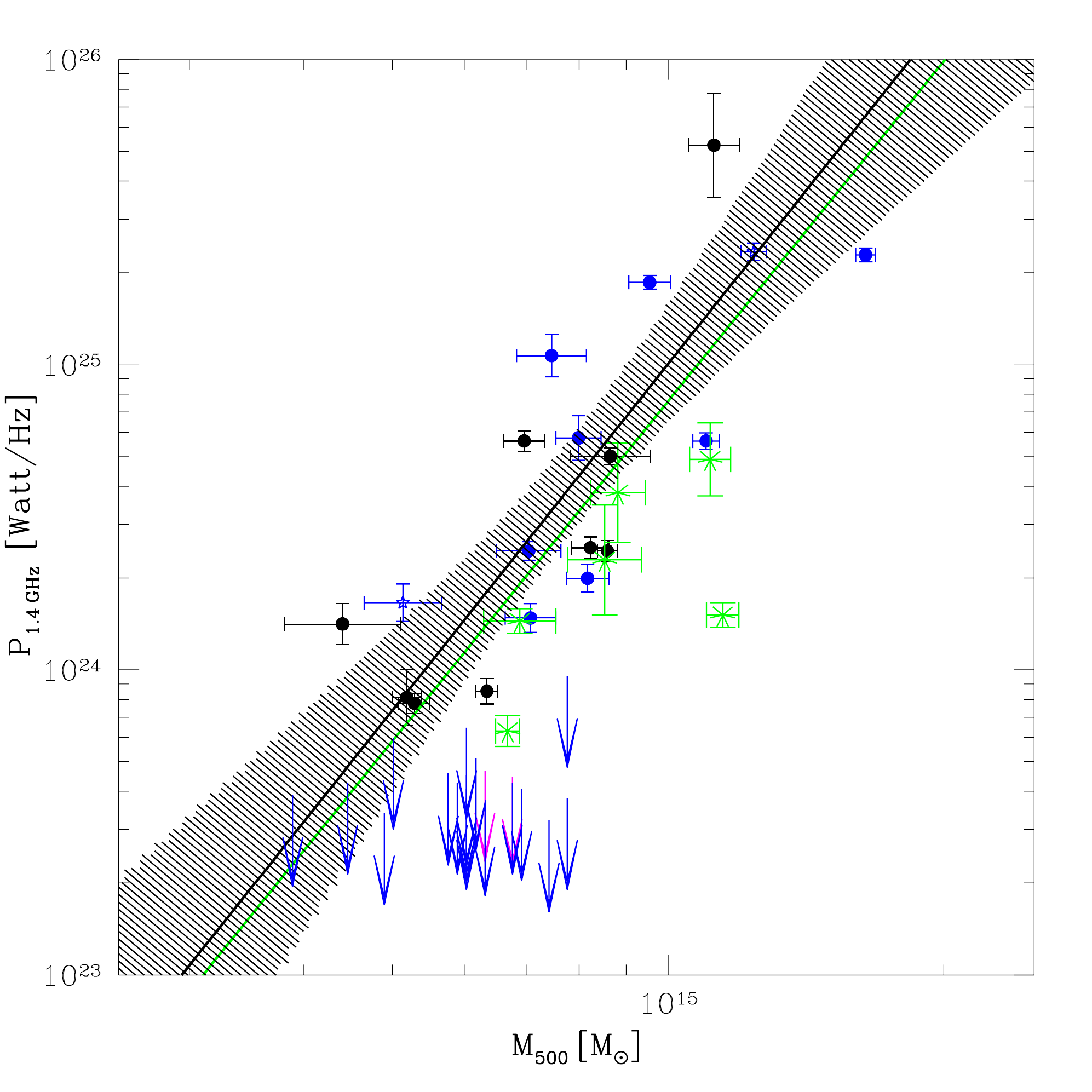}
\includegraphics[width=0.45\textwidth]{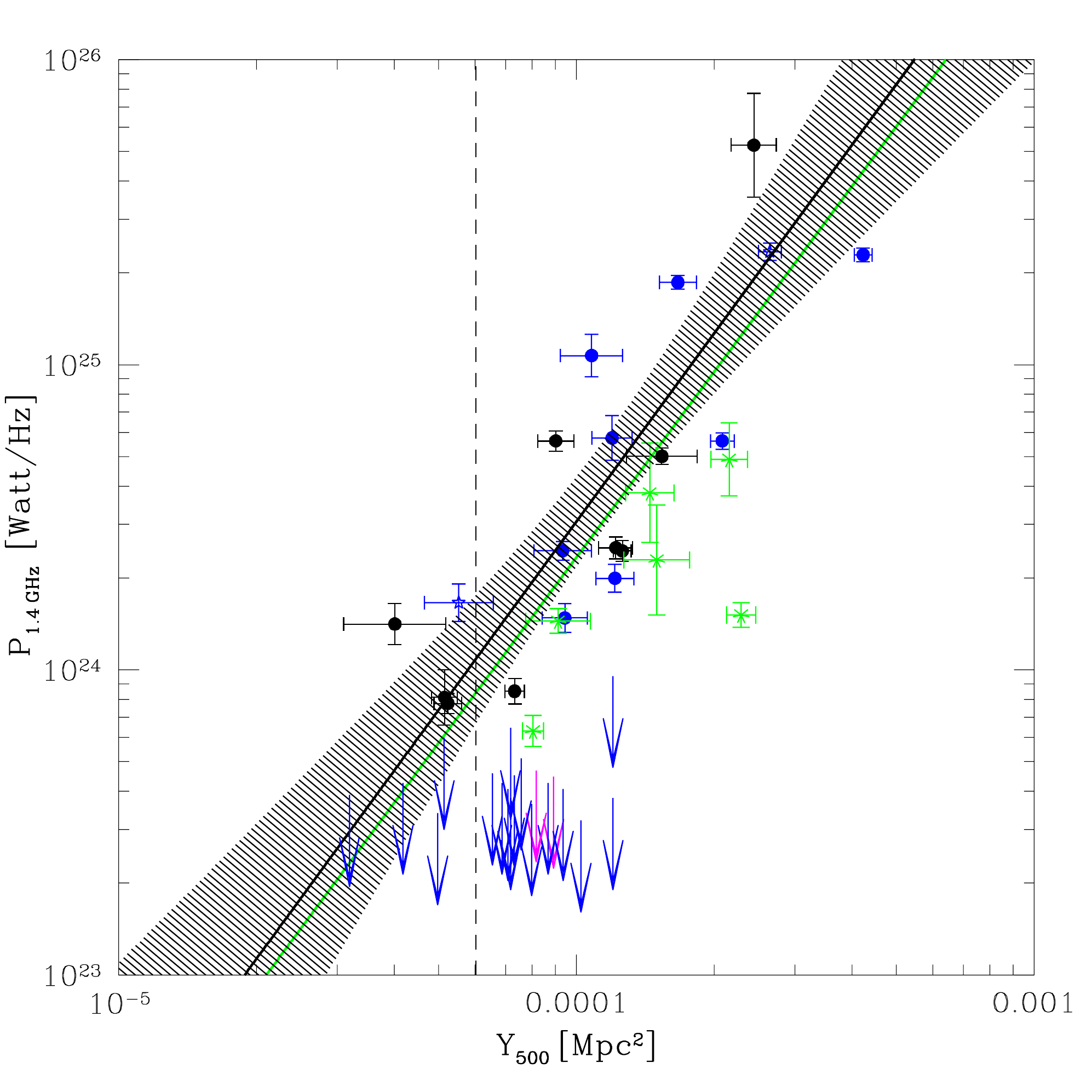}
\caption[]{Distribution of clusters in the $P_{1.4} - M_{500}$ ({\it left panel}) and in the $P_{1.4} - Y_{500}$ diagrams ({\it right panel}). In both panels different symbols are as in 
Fig.~\ref{Fig.Lr_Lx}. Best-fit relations to giant RHs only (black lines) and to all RHs (including USSRH, green dashed lines) are reported. Dashed line in the {\it right panel} marks the value $Y_{500}=6\times10^{-5}$ Mpc$^2$.}
\label{Fig.Lr_Y}
\end{center}
\end{figure*}

\section{Radio-SZ scaling relations}

As discussed in Sect.\ref{Sect.SZ}, observations of clusters through their SZ-effect may provide a powerful method to measure the cluster masses. Recently, Basu (2012) found a correlation between the radio power of clusters with RHs and the integrated Compton parameter derived from the {\it Planck} ESZ catalogue (Planck Collaboration 2011a) in the form $P_{1.4}\propto Y_{5R_{500}}^2$, where $Y_{5R_{500}}$ is the integral of the SZ signal within a radius of $5R_{500}$\footnote{$Y_{5R_{500}}$ can be rescaled to $Y_{500}$ for the fiducial GNFW model as $Y_{5R_{500}}=1.79\times Y_{500}$ (Arnaud et al. 2010).}. Basu (2012) found indication for a weaker or lack of bimodality based on the fact that only 4 clusters from the GRHS with radio upper limits were found in the {\it Planck} ESZ catalogue, while almost all RH of the GRHS have counterparts in the same catalogue. Basu (2012) suggested that a possible reason for the lack of bimodality in SZ could be due to the fact that X-ray selected cluster samples are biased towards the detection of X-ray luminous, but not necessary massive, clusters, while the SZ tends to be more ``mass-limited". In this picture, clusters with radio upper limits that are not detected by {\it Planck} should be less massive systems (with respect to those hosting giant RHs) with cool-cores. These clusters would appear brighter in X-ray because of the presence of a cool core, causing an apparent bimodality in the $P_{1.4}-L_{500}$ plane. However, as we have shown in Sect.\ref{sec:radioX}, even when we consider the X-ray luminosity excising the cool core, we find a clear bimodality in the radio-X-ray plane (Fig.~\ref{Fig.Lr_Lx}, {\it right panel}). 

The all-sky PSZ catalogue, that we are using in this paper, is six times the size of the {\it Planck} ESZ catalog (Planck Collaboration 2013b) used by Basu (2012), and $\sim80\%$ complete for $M_{500}\gtsim 6\times10^{14}\,M_{\odot}$ at $z\simeq 0.2-0.35$, typical mass and redshift ranges of the EGRHS clusters. In Fig.~\ref{Fig.Lr_Y}, we show the distribution of the 44 clusters of our sample belonging to the PSZ catalogue (see Sect.\ref{Sect.SZ}), in the $P_{1.4}-M_{500}$ ({\it left panel}) and $P_{1.4}-Y_{500}$ ({\it right panel}) diagrams. We show with different colors clusters belonging to the EGRHS (blue points and blue and magenta arrows), halos from the literature (black points) and halos with ultra-steep radio spectra (green asterisks).The comparison between RHs and upper limits can be performed only for clusters belonging to the EGRHS, while the RHs from the literature are added to better determine the correlations.
We find clear correlations between $P_{1.4}$ and $M_{500}$ and $Y_{500}$ parameters. Using the BCES regression method, we fit the observed $P_{1.4}-Y_{500}$ and $P_{1.4}-M_{500}$ relation with the following power laws:

\begin{equation}
\log\Big(\frac{P_{1.4}}{10^{24.5}\mathrm{Watt/Hz}}\Big)=B\,\log\Big(\frac{Y_{500}}{10^{-4}\mathrm{Mpc^2}}\Big)+A
\end{equation}

\noindent and

\begin{equation}
\log\Big(\frac{P_{1.4}}{10^{24.5}\mathrm{Watt/Hz}}\Big)=B\,\log\Big(\frac{M_{500}}{10^{14.9}\mathrm{M_{\odot}}}\Big)+A
\end{equation}

\begin{figure}
\begin{center}
\epsscale{1.0}
\plotone{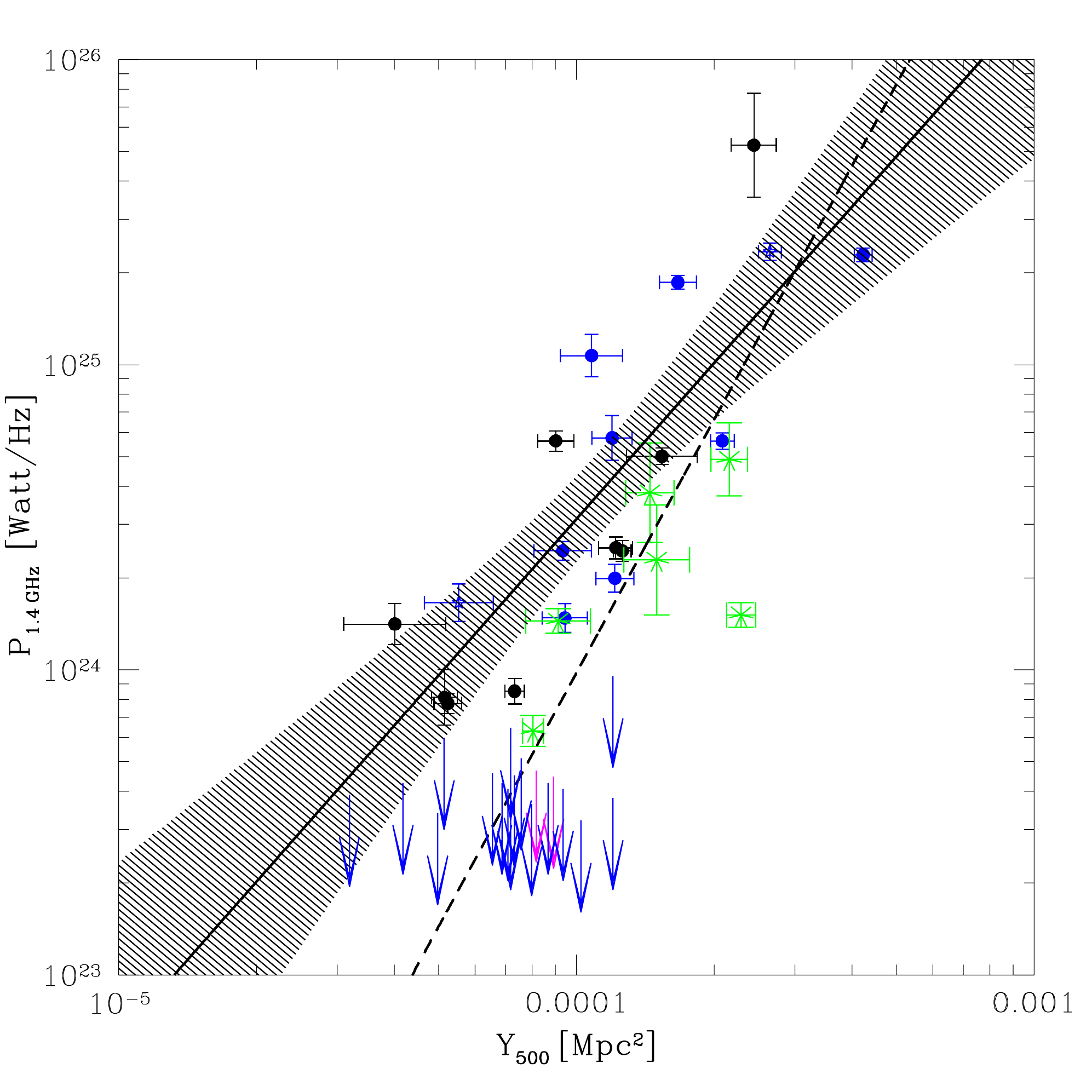}
\caption[]{Distribution of clusters in $P_{1.4} - Y_{500}$ plane. Symbols are as in Fig.~\ref{Fig.Lr_Lx}. Best-fit relations to giant RHs (black solid line) and to giant RHs plus upper limits (dashed line) are also shown. The shadowed region show the 95\% confidence region of the best-fit correlation for giant RHs.}
\label{Fig.Lr_Y_95}
\end{center}
\end{figure}
							    
\begin{figure}
\begin{center}
\epsscale{1.0}
\vskip 0.5 mm
\plotone{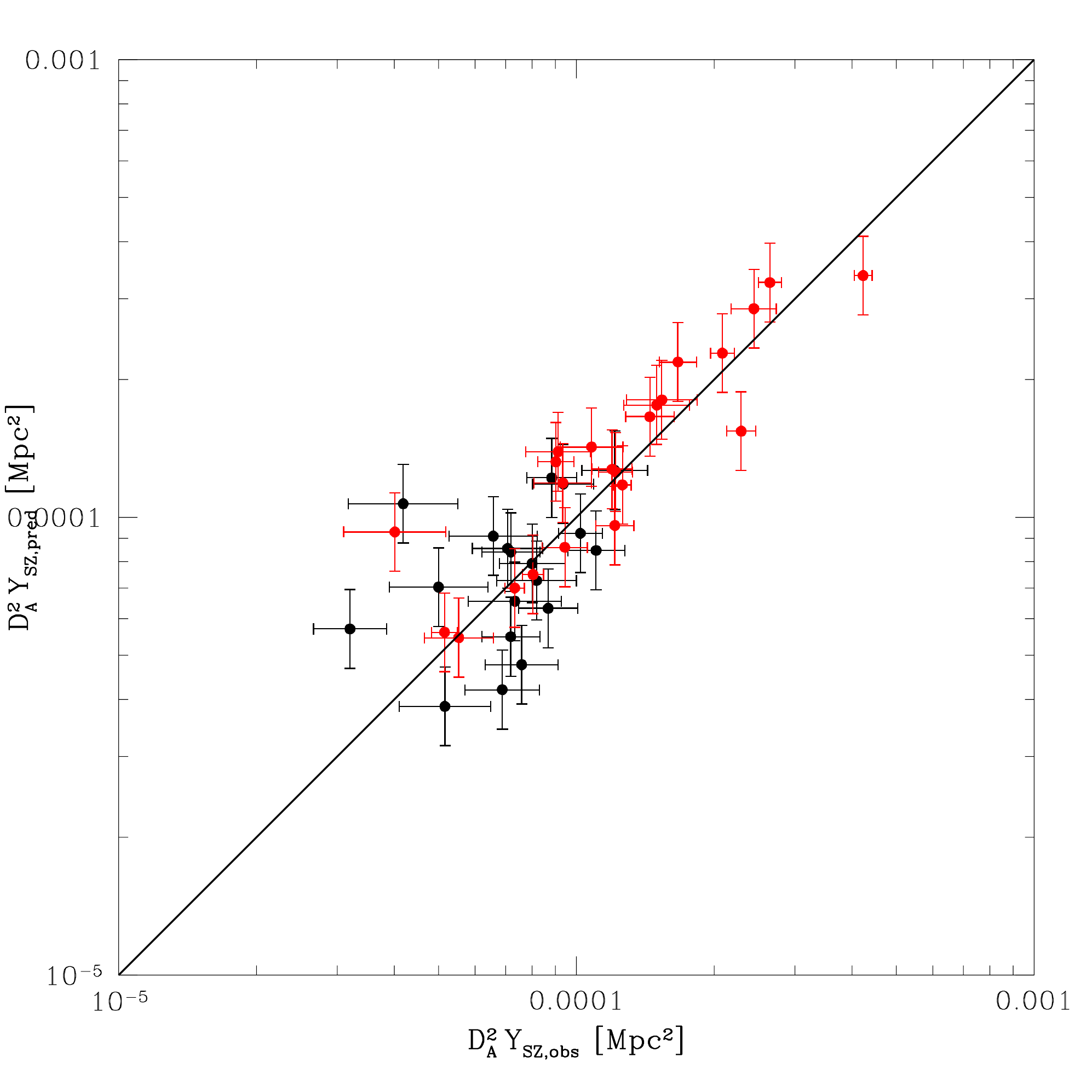}
\caption[]{Comparison between the observed values of $Y_{500}$ (in abscissa) and those predicted by the $Y_{500}-L_{500,nc}$ relation, $Y_{500,pred}$ (in ordinate) for RH clusters (red points) and clusters with radio upper limits (black points). The black solid line shows the one to one trend.}
\label{Fig.YSZoYSZd}
\end{center}
\end{figure}

Results of the fits, together with those from 1000 bootstrap resamples, are reported in Table~\ref{Tab.bestfit}. The slope of the $P_{1.4}-Y_{500}$ correlation is close to $\sim 2$, consistent with that found by Basu (2012); it is $2.05\pm0.28$ when the BCES-bisector method is used, and $2.28\pm0.35$ when the BCES-orthogonal method is adopted.
The slope of the $P_{1.4}-M_{500}$ correlation is $3.77\pm0.57$ and $4.51\pm0.78$ in the case of the BCES-bisector and BCES-orthogonal methods, respectively, both steeper than previous estimates, because of the different definitions of the cluster masses (within a fixed size of 3 Mpc, Feretti 2003; or the virial mass, Cassano et al. 2006).

At variance with Basu (2012) we find a clear bimodal behavior of clusters in both diagrams. For $M_{500}\gtsim 5.5\times 10^{14}\,M_{\odot}$ and for $Y_{500}\gtsim 6\times 10^{-5}$ Mpc$^{2}$, all clusters with radio upper limits are well below the 95\% confidence region of the best-fit correlations.
For the sake of completeness, for the $P_{1.4}-Y_{SZ}$ relation, we also performed a regression analysis by making use of the parametric EM algorithm that also deals with upper limits (see Sect.\ref{Sect.method}). This allows to evaluate the effect of the radio upper limits on the best-fit correlation, and thus to test the reliability of the correlation and the presence of a bimodal behavior in the cluster radio powers. The best-fit values are reported in Table~\ref{Tab.bestfit} and the best-fit correlations obtained for giant RHs only and for giant RHs plus upper limits are shown in Fig.~\ref{Fig.Lr_Y_95} (solid and dashed line, respectively) together with the 95\% confidence region of the RH-only correlation. All upper limits (with just one exception) lie below the 95\% confidence region, and the two best-fit relations obtained by considering RHs plus upper limits or only RHs differ both in slope and in normalization.

Our statistical analysis suggests two distinct populations of clusters: those with giant RHs, occupying the region of the correlation, and those without giant RHs, separated from that region. 

\begin{figure}
\begin{center}
\epsscale{1.0}
\plotone{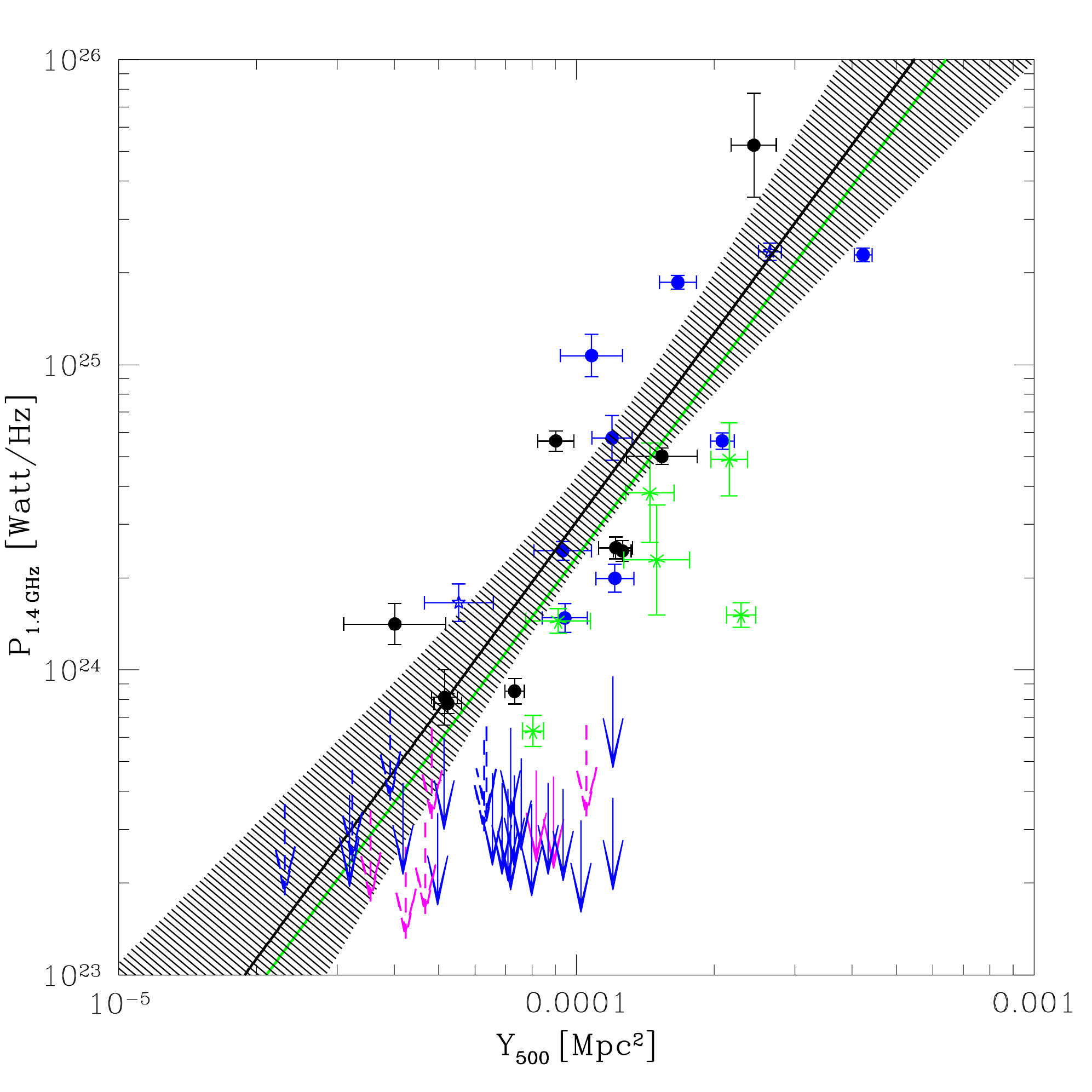}
\caption[]{Distribution of clusters in the plane $P_{1.4} -Y_{500}$. Symbols are like in Fig.~\ref{Fig.Lr_Lx}, with dashed arrows indicating the predicted positions of clusters currently not present in the {\it Planck} catalogue. The best-fit to giant RHs only (black solid line) and to giant RHs plus USSRH (green line) are also shown. The shadowed region show the 95\% confidence region of the best-fit correlation for giant RHs only.}
\label{Fig.LrYSZ}
\end{center}
\end{figure}

\subsection{Non-detected {\it Planck} clusters in the $P_{1.4}-Y_{500}$.}

With the aim to evaluate the possible position, in the $P_{1.4}-Y_{500}$ diagram, of EGRHS clusters not contained in the 15.5 months PSZ validation catalogue, we make use of the correlation between $Y_{500}$ and the core-excluded X-ray luminosity, $L_{500,nc}$ (Sect.\ref{Sect.Lx}). By using the {\it Planck-XMM-Newton} archive sample, which comprises 62 clusters with the highest quality X-ray and SZ data set currently available (Planck Collaboration 2011b), we derive the $0.1-2.4$ keV X-ray luminosity between $[0.15-1]R_{500}$ ($L_{500,nc}$, hereafter) and obtained the following $Y_{500}-L_{500,nc}$ correlation:

\begin{equation}
h(z)^{-2/3} Y_{500} = A\Big(\frac{h(z)^{-7/3}\,L_{\mathrm{500,nc}}}{7\times 10^{44}\mathrm{erg/s}}\Big)^B\mathrm{Mpc^2}
\label{Eq.YSZ}
\end{equation}

\noindent where $h(z)=\sqrt{\Omega_m(1+z)^3+\Omega_{\Lambda}}$, $A=10^{-3.795\pm0.014}$ and $B=1.094\pm0.039$. We thus derive $L_{500,nc}$ for all clusters in Table~\ref{Tab.ULRH}\footnote{with the exceptions of Coma, MACS1149.5+2223 and PLK171.9-40.7, for which the $L_{500}$ were taken from the literature (see Table~\ref{Tab.ULRH}).} and then apply Eq.~\ref{Eq.YSZ} to estimate their $Y_{500}$ parameters. To test the consistency of this approach, we compare the ``observed'' and ``predicted'' values of $Y_{500}$ for the clusters present in the PSZ catalogue. Such a comparison is shown in Fig.~\ref{Fig.YSZoYSZd}: the data are consistent with a one-to-one trend (with increasing scatter at lower values), suggesting that indeed we can apply this procedure to get reliable estimates of the $Y_{500}$ for clusters not contained in the PSZ catalogue.

In Fig.~\ref{Fig.LrYSZ}, we show the distribution of all clusters in the $L_{1.4} - Y_{500}$ diagram, including those that are actually not observed by {\it Planck} (dashed arrows). 
As expected, the bulk of clusters missing in the PSZ catalogue is in the region of lower $Y_{500}$ values and with $M_{500}<5.5\times 10^{14}\,M_{\odot}$, where the PSZ catalogue
is only marginally complete (the completeness is $\sim$20\%). There are however two exceptions : RXCJ1532.9+3021, a luminous cool core cluster, and RXCJ0437.1+0043, which are expected in the region of massive clusters.

\begin{figure*}
\begin{center}
\epsscale{1.0}
\plottwo{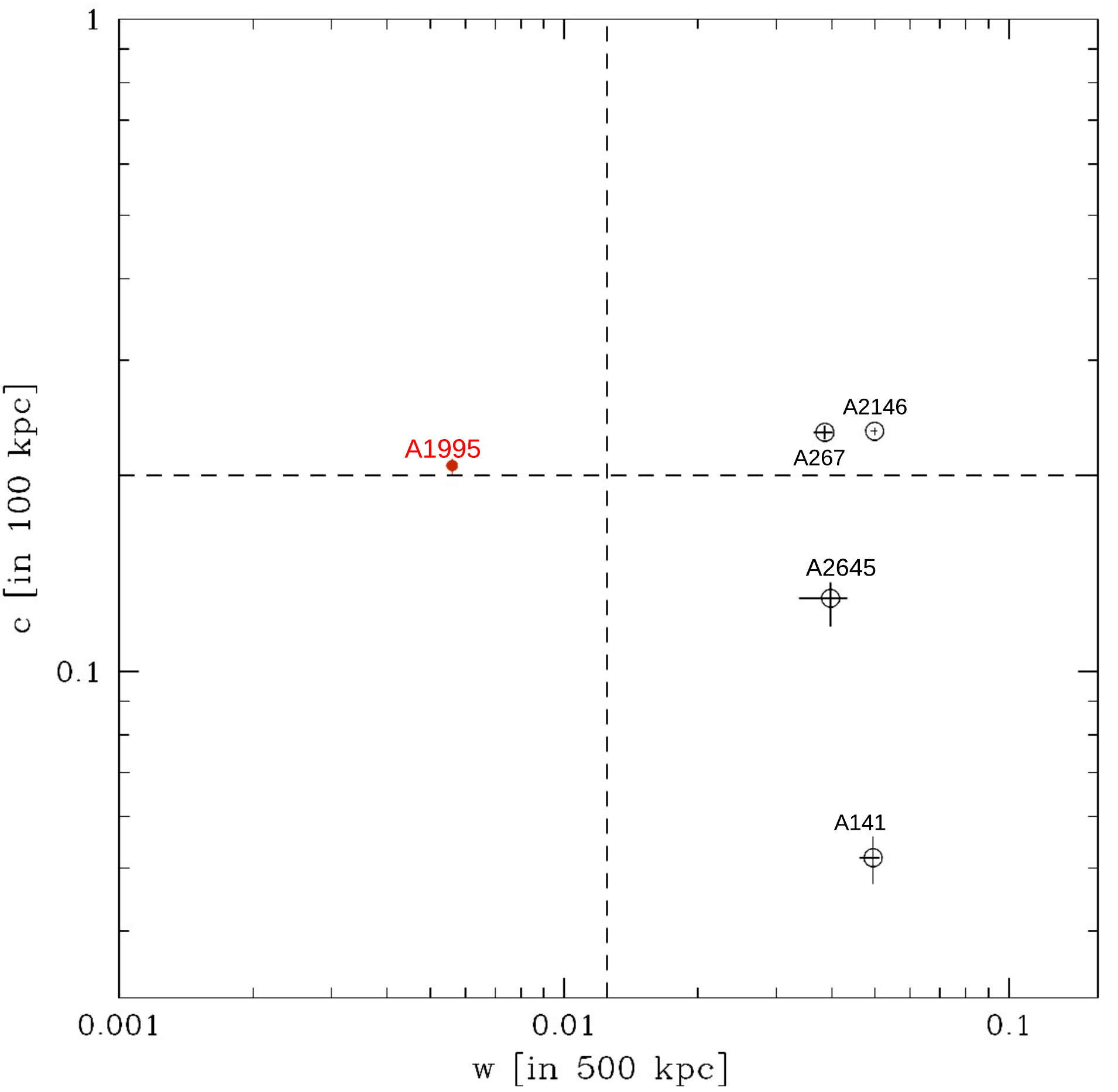}{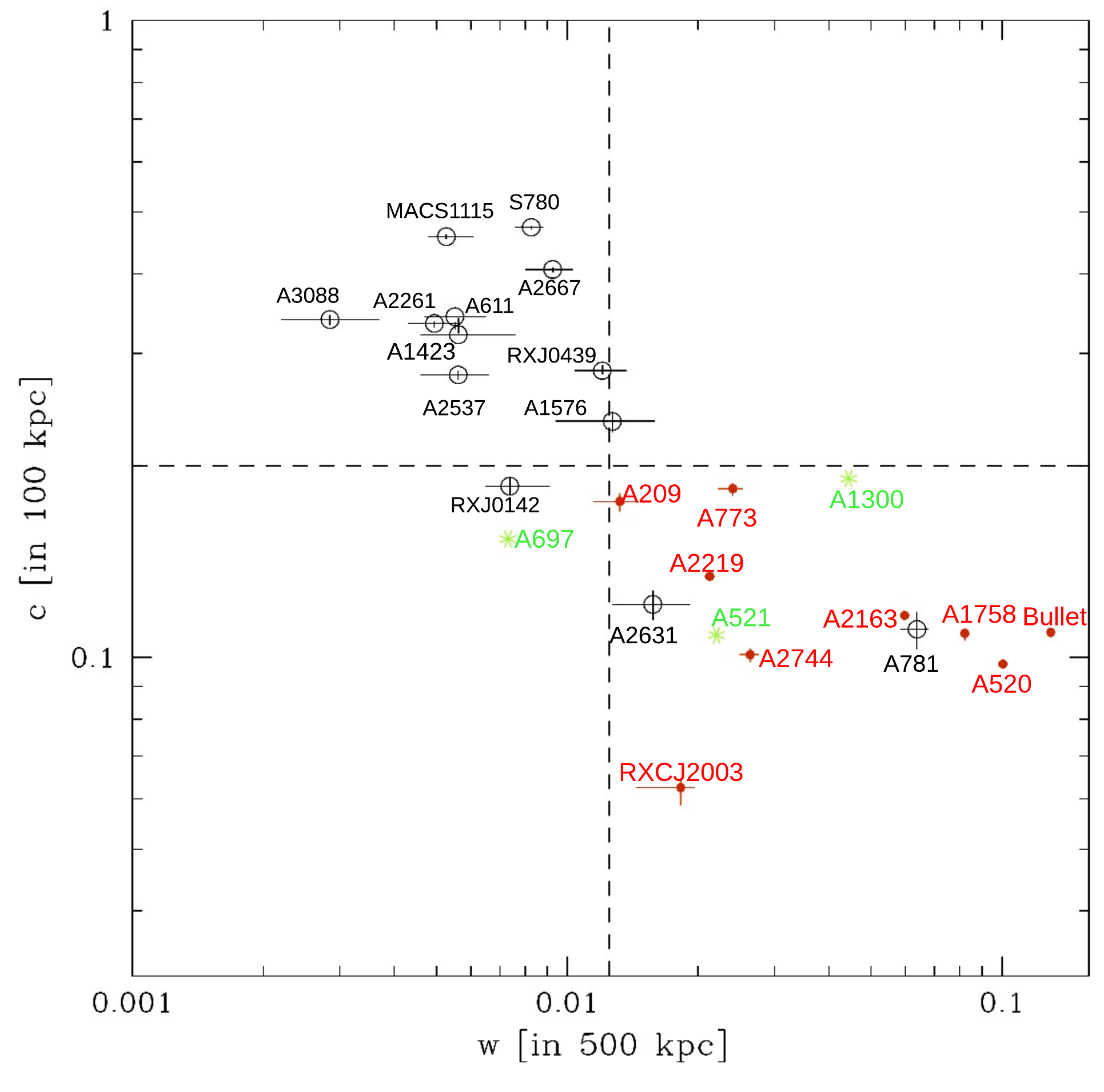}
\caption[]{Distribution of clusters in the plane $c-w$.  Clusters with $Y_{500}\ltsim 6\times 10^{-5}\,{\mathrm Mpc^2}$ ({\it left panel}) and clusters with $Y_{500}\gtsim 6\times 10^{-5}\,{\mathrm Mpc^2}$ ({\it right panel}) are reported. Black open points are clusters with radio upper limits, while clusters with giant RH and with USSRH are shown as red points and green asterisks, respectively. Vertical and horizontal dashed lines: $c=0.2$ and  $w=0.012$.}
\label{Fig.dynamics}
\end{center}
\end{figure*}

\subsection{On the $P_{1.4}-Y_{500}$ scaling relation}

If we focus on clusters belonging to the EGRHS and also consider the two clusters (A1995 and the {\it Bullet} cluster) which are in the same X-ray luminosity and redshift range of the EGRHS clusters\footnote{A1995 belong to the NORAS survey which has a slightly lower flux limit with respect to the eBCS used to select the GRHS; the {\it Bullet} cluster is in the south and not easily accessible for the {\it GMRT}.},
we find a segregation of clusters in the $P_{1.4}-Y_{500}$ diagram for $Y_{500}\gtsim 6\times10^{-5}$ Mpc$^2$: clusters with RHs follow a trend between their radio power and the cluster SZ parameter, while clusters without RHs populate the region of radio upper limits, which is a factor of $\sim 5-7$ below the correlation (Fig.~\ref{Fig.Lr_Y}, {\it right panel}). On the other hand, for $Y_{500}\ltsim 6\times10^{-5}\, \mathrm{Mpc^2}$, upper limits are not deep enough and lie within the 95\% confident region of the best-fit correlation.

In order to better understand this behavior of clusters and shed light on the mechanism responsible for the formation of giant RHs in clusters, we looked at the dynamical properties of clusters in the $P_{1.4}-Y_{500}$ diagram, adopting  the centroid shift variance, $w$, and the surface brightness concentration parameter, $c$, to differentiate between merging 
($w<0.012$ and $c>0.2$) and more relaxed ($w>0.012$ and $c<0.2$) systems (see Sect.\ref{Sect.dyn}).

For $Y_{500}\ltsim 6\times10^{-5}\, \mathrm{Mpc^2}$, we find that four clusters with radio upper limits detected by {\it Planck} are merging clusters (Fig.~\ref{Fig.dynamics}, left panel). Unfortunately, the $L_{1.4}-Y_{500}$ correlation predicts a RH power for those clusters that is below the sensitivity of the current radio data.
In light of our results, the apparent lack of a giant RH in the merging but relatively low-mass cluster Abell 2146 (Russell et al. 2011) is not surprising, because even if a halo is present in this cluster, it may not be luminous enough to be detected.
The only RH cluster with $Y_{500}\ltsim 6\times10^{-5}\, \mathrm{Mpc^2}$ is A1995, which in Fig.~\ref{Fig.dynamics} (left panel) is located in a region generally populated by ``relaxed'' clusters. However, A1995 is a merging system, but the merger is happening mainly along the line of sight (Boschin, Girardi \& Barrera 2012), and for this reason its position in the $c-w$ diagram is likely biased by projection effects.

Clusters with $Y_{500}\gtsim 6\times10^{-5}\, \mathrm{Mpc^2}$ show a clear segregation in their dynamical properties. All clusters with detected giant RHs are clearly merging systems, while the majority of clusters with upper limits ($\sim 80\%$) are more relaxed (Fig.~\ref{Fig.dynamics}, right panel). The presence of a segregation in the dynamical state of clusters with detected and non-detected RH strengthens the separation of clusters in the $P_{1.4}-Y_{500}$ diagram and suggests that mergers have a crucial role in the formation of these cluster-wide diffuse radio sources.

Another interesting observations is that all clusters with $Y_{500}>1.3\times10^{-4}\,\mathrm{Mpc^2}$ are merging clusters and host a giant RH. These clusters are very massive systems with $M_{500}\gtsim8\times 10^{14}\,M_{\odot}$. In particular, if we consider only clusters belonging to the EGRHS (plus the ``Bullet'' cluster lying within the same redshift range), we have 6 clusters with $Y_{500}>1.3\times10^{-4}\,\mathrm{Mpc^2}$: 4 giant RHs and 2 USSRH. Why do we not find massive relaxed clusters in the EGRHS? The EGRHS is an X-ray selected sample, thus there are no reasons why we should miss a population of massive relaxed clusters, which are generally X-ray luminous. A possibility is that the $Y_{500}$ estimates for merging clusters are biased high with respect to $M_{500}$.
Numerical simulations show that merging clusters fall below the $M-Y$ scaling relation, such that their inferred masses could be biased low (\eg Krause et al. 2012). However, recent observations based on SZ and weak-lensing cluster mass measurements show that merging clusters have weak-lensing masses 40\% lower than relaxed clusters at fixed $Y_{500}$, so that their inferred SZ masses are biased high (\eg Marrone et al. 2012). The latter authors suggested that the possible cause of these discrepancies could be found in the over-simplicity of the adopted models to fit the weak-lensing data.

\noindent A more promising hypothesis is that the lack of massive relaxed systems in the EGRHS is due to the redshift range of this sample,  $z\simeq0.2-0.4$, that is not far from the formation epoch of these massive systems, $M_{500}\gtsim 8 \times 10^{14}\,M_{\odot}$  (\eg Giocoli et al. 2007, 2012). In this case the probability to observe massive relaxed clusters is smaller; we will investigate these points in more detail in a separate paper (Cassano et al., in prep.).

\section{Summary \& Conclusions}

A number of correlations between thermal and non-thermal cluster properties, \ie $P_{1.4}-L_X$, $P_{1.4}-M$ and $P_{1.4}-T_X$, have been reported for clusters hosting giant RHs since the last decade. However, due to the small statistics and to the lack of statistical samples of clusters observed at radio wavelengths,
the reliability of these correlations and the effects of observational biases were not clear (\eg Rudnick et al. 2006).
Only recently, thanks to the GRHS (Venturi et al. 2007, 2008) it has been possible to rely upon a solid sample of clusters with homogeneous and deep radio observations. For the first time, it was possible to place firm upper limits to the diffuse radio flux of clusters without extended diffuse radio emission at the detection level of the survey. These upper limits allowed for the study of the distribution of clusters in the $P_{1.4}-L_X$ and to discover a bimodal behavior in the population of clusters: RH clusters lying on the $P_{1.4}-L_X$ correlation and radio-quiet clusters (Brunetti et al. 2007; 2009). Most important, the separation between RH  and radio-quiet clusters has a correspondence in the dynamical state of clusters, with merging systems that harbor RHs and radio-quiet clusters that are statistically more relaxed (Cassano et al. 2010).
The bimodality has been questioned in the light of the cross-correlation of the GRHS with the {\it Planck} ESZ cluster catalogue. It was  
shown that while almost all RHs have been detected in SZ, only 4 out of 20 upper limits were detected (Basu 2012). 
This was interpreted as a weaker or absent bimodality in the radio-SZ plane. The proposed explanation for this was that SZ measurements allow an unbiased estimate of the cluster mass, whereas X-ray based cluster samples are biased towards the detection of bright cool core clusters, that may induce an apparent bimodal distribution of clusters in the radio-X-ray plane (Basu 2012).

In this paper, we revise the radio-X-ray and radio-SZ correlations. Our analysis is based on the EGRHS (Kale et al. 2013). We searched and found information in the {\it ROSAT} and {\it Chandra} archive for a sub-sample of 40 clusters: 29 with upper limits to the radio powers and 11 with giant RHs. In addition to this sample, we also found information for a sample of 14 clusters hosting well-known RHs from the literature. These are used to obtain a better leverage in radio/X-ray luminosities, which helps in the derivation of more robust scaling relations. 

First, we derive the correlation between the monochromatic radio power of halos at 1.4 GHz and the 0.1-2.4 keV band X-ray luminosity of the parent cluster. We revaluate in a homogeneous way the radio flux of all the halos by using {\it GMRT} and literature data and measure the X-ray luminosity of the clusters within $R_{500}$ from pointed {\it ROSAT} observations  and {\it Chandra} when {\it ROSAT} data are not available (or not sensitive enough). 
For the first time we show the presence of a scaling $P_{1.4\,{\mathrm GHz}}\propto L_{500}^{2.1\pm0.2}$. Being steeper than other halos, USSRH are in general under-luminous with respect to this correlation. Their inclusion in the fit procedure produces a slightly lower normalization and an increase of  the scatter with respect to the best-fit relation. We also correct the X-ray luminosity of the parent cluster by modeling the X-ray brightness distribution and excising the cool core. We find that for $L_{500}$ (or $L_{500,cor}$)$\gtsim 5\times 10^{44}$ erg/sec the distribution of clusters in the ($P_{1.4\,{\mathrm GHz}}, L_{500}$) and ($P_{1.4\,{\mathrm GHz}}, L_{500,cor}$) planes is bimodal: RH clusters lie on the correlation, while clusters with upper limits to the radio power are below the 95\% confidence region of the best-fit correlation. This allows to conclude that the presence of cool-core clusters does not affect the bimodal behavior of clusters in the radio power X-ray luminosity plane. 

To investigate the behavior of clusters in the radio-SZ diagram, we cross-checked the sample of clusters selected from the EGRHS with the 15.5 month {\it Planck} SZ catalogue (PSZ; Planck Collaboration 2013b) and found SZ information for all 11 RHs and for 19 out of 29 clusters with upper limits. Also for the remaining 14 clusters with giant RHs we found information in the PSZ catalogue. We found a clear correlation between the RH $P_{1.4}$ and the cluster $Y_{500}$ of the form $P_{1.4}\propto Y_{500}^{2.05\pm0.28}$, in line with previous findings (Basu 2012). However, contrary to previous findings, at least for $Y_{500}\gtsim 6\times10^{-5}\, {\mathrm Mpc^2}$ (roughly corresponding to $M_{500}\gtsim 5.5\times 10^{14}\,M_{\odot}$) we find that all clusters with radio upper limits lie below the 95\% confidence region of the best-fit correlation, highlighting a bimodal behavior of clusters in the radio-SZ diagram. This segregation is strengthened by the separation of those clusters in the morphological diagrams: clusters with diffuse radio emission are merging clusters, while the great majority of clusters with upper limits are relaxed, thus highlighting the importance of merging events in the generation of giant RHs.
We also use the tight correlation between the core-excised cluster X-ray luminosity $L_{500,nc}$ and $Y_{500}$ to derive the predicted value of $Y_{500}$ for those clusters in our sample that are actually not detected by {\it Planck}. As expected, we found that the majority of them (8 out of 10) are in clusters with $Y_{500}\ltsim 6\times10^{-5}\, {\mathrm Mpc^2}$, where the completeness of the PSZ catalogue is poor (about 20\%). Interestingly, half of the non-detected clusters are cool core clusters, only two of seven cool core clusters of our sample were detected by {\it Planck}, suggesting that in the region of lower completeness {\it Planck} loses preferentially cool-core clusters with respect to merging systems.

The EGRHS is not selected in mass but in X-ray luminosity. However, considering that the completeness of the PSZ catalogue for $M_{500}\geq 6\times10^{14}\,M_{\odot}$ at $0.2\leq z\leq0.33$ is $\sim 80\%$ (Planck Collaboration 2013b) and cross-correlating the PSZ catalogue with the EGRHS, we estimated that the completeness in mass of the EGRHS is $\sim55\%$, and the addition of radio observations of $\sim 17$ galaxy clusters from the PSZ catalogue will provide a sample of mass selected clusters with deep radio data and a completeness of $\sim 80\%$. For a comparison, assuming the same masses and redshift range, we estimated that the completeness of the ESZ {\it Planck} catalogue is of the order of 35\%.

Remarkably, we found that for $Y_{500}\gtsim 1.3\times10^{-4}\, {\mathrm Mpc^2}$ (or $M_{500}\gtsim 8\times 10^{14}\,M_{\odot}$) all clusters of the EGRHS are in the process of merging and have a RH. We consider several possibilities to explain this result and conclude that the most likely explanation is that we are looking at these massive systems near their formation epoch (we selected clusters at $z\sim 0.2-0.4$) and thus the probability to observe massive relaxed systems at these redshift should be relatively low.

In Sect.4 we derive basic scaling relations predicted by the two main scenarios put forward so far to explain giant RH.  Under the assumption that synchrotron emission dominates energy losses of relativistic electrons in the ICM, and that the ratio between the energy density of cosmic-ray protons and thermal ICM in the radio emitting volume does not depend on cluster mass, ``secondary models'' predict hat the synchrotron power of the halos scales as $\nu P^{syn}\propto L_X^{1.6-1.5}$ (\eg Kushnir et al. 2009) and $\nu P^{syn}\propto Y_{500}^{1.55-1.43}$. These scalings are flatter than those derived from observations in the present paper (see Tab.\ref{Tab.bestfit}). Re-acceleration models typically predict steeper slopes. For example, under similar assumptions for magnetic field and cosmic rays, following Cassano et al. (2007), the scalings of the halo radio power with the cluster mass and SZ flux are $\nu P^{syn}\propto M_{500}^{4} $ and $\nu P^{syn}\propto Y_{500}^{2.3}$, respectively; and are in agreement with the observed scalings (see Tab.\ref{Tab.bestfit}). A detailed comparison between model expectations and observed scalings which considers the full range of model parameters is beyond the aim of this paper.

It is also worth mentioning that in both the radio-X-ray and radio-SZ diagrams clusters with USSRH are all below the 95\% confidence region of the best-fit correlations. They are preferentially located in the region between ``classical'' RHs and radio upper limits. This is not surprising, since these RHs are 
steeper than those on the correlations and thus their synchrotron emissivity at 1.4 GHz is lower with respect to that of RHs with flatter spectra. Interestingly, their position relative to the correlations was already predicted by models in which RHs are generated as a result of the turbulent re-acceleration of relativistic electrons in the ICM (\eg Cassano 2010; Donnert et al. 2013).

\section{Acknowledgments}
We thank the referee for the useful comments. RC thanks Nemmen Rodrigo for providing the patched BCES routine running with gfortran in Mac, and N. Aghanim, D. Dallacasa, J. Donnert, C. Giocoli, P. Mazzotta and F. Vazza, for useful discussions.  We thank H. Bourdin for supplying additional data. K.D. acknowledges the support by the DFG Cluster of Excellence ``Origin and Structure of the Universe''. S.G. acknowledges the support of NASA through Einstein Postdoctoral Fellowship PF0-110071 awarded by the {\it Chandra} X-ray Center (CXC), which is operated by SAO.


\begin{thebibliography}{}


\bibitem[Ackermann et al.(2010)]{2010ApJ...717L..71A} Ackermann, M., Ajello, M., Allafort, A., et al.\ 2010, \apjl, 717, L71 
\bibitem[Akritas \& Bershady(1996)]{Akritas96} Akritas M.G.,Bershady M.A., 1996, ApJ 470, 706
\bibitem[Arnaud \& Evrard(1999)]{1999MNRAS.305..631A} Arnaud, M., \& Evrard, A.~E.\ 1999, \mnras, 305, 631
\bibitem[Arnaud et al.(2005)]{2005A&A...441..893A} Arnaud, M., Pointecouteau, E., \& Pratt, G.~W.\ 2005, \aap, 441, 893
\bibitem[Arnaud et al.(2010)]{2010A&A...517A..92A} Arnaud, M., Pratt, G.~W., Piffaretti, R., et al.\ 2010, \aap, 517, A92

\bibitem[Basu(2012)]{2012MNRAS.421L.112B} Basu, K.\ 2012, \mnras, 421, L112

\bibitem[Biffi et al.(2013)]{2013MNRAS.428.1395B} Biffi, V., Dolag, K., B{\"o}hringer, H.,\ 2013, \mnras, 428, 1395 

\bibitem{}B{\"o}hringer, H., Schuecker, P., Guzzo, L., et al., 2004, A\&A, 425, 367
\bibitem{}B\"ohringer, H.; Pratt, G. W.; Arnaud, M.; et al., 2010, A\&A, 514, 32

\bibitem[Boschin et al.(2012)]{2012A&A...547A..44B} Boschin, W., Girardi, M., \& Barrena, R.\ 2012, \aap, 547, A44


\bibitem[Bonafede et al.(2009)]{2009A&A...503..707B} Bonafede, A., Feretti, L., Giovannini, G., et al.\ 2009, \aap, 503, 707
\bibitem[Bonafede et al.(2012)]{2012MNRAS.426...40B} Bonafede, A., Br{\"u}ggen, M., van Weeren, R., et al.\ 2012, \mnras, 426, 40 

\bibitem[Brown et al.(2011)]{2011ApJ...740L..28B} Brown, S., Emerick, A., Rudnick, L., \& Brunetti, G.\ 2011, \apjl, 740, L28
\bibitem[Brunetti et al.(2001)]{Bru01a} Brunetti G., Setti G., Feretti L., Giovannini G., 2001, MNRAS 320, 365
\bibitem{}Brunetti, G., Venturi, T., Dallacasa, D., et al., 2007, ApJ, 670L, 5B
\bibitem{}Brunetti, G., Giacintucci, S., Cassano, R., et al. , 2008, Nature, 455, 944 
\bibitem{}Brunetti, G., Cassano, R., Dolag, K., Setti, G., 2009, A\&A 507, 661
\bibitem[Brunetti et al.(2012)]{2012MNRAS.426..956B} Brunetti, G., Blasi, P., Reimer, O., et al.\ 2012, \mnras, 426, 956

\bibitem{}Buote, D.A., 2001, ApJ, 553, L15
\bibitem[Colafrancesco(1999)]{Cola99} Colafrancesco, S.\ 1999, ``Diffuse Thermal and Relativistic Plasma in Galaxy Clusters'', 269; :astro-ph/9907329

\bibitem{}Cassano, R., 2009, in {\it The Low Frequency Radio Universe}, ASP Conf.Ser., Vol. 407, 223, Eds. D.J. Saikia, D. Green, Y. Gupta \& T. Venturi
\bibitem{}Cassano, R., 2010, A\&A 517, 10
\bibitem[Cassano et al.(2006)]{2006MNRAS.369.1577C} Cassano, R., Brunetti, G., \& Setti, G.\ 2006, \mnras, 369, 1577
\bibitem{}Cassano, R., Brunetti, G., Setti, G., et al. , 2007, MNRAS, 378, 1565
\bibitem{}Cassano, R., Brunetti, G., Venturi, T., et al., 2008, A\&A, 480, 687 
\bibitem[Cassano et al.(2010)]{2010ApJ...721L..82C} Cassano, R., Ettori, S., Giacintucci, S., et al.\ 2010, \apjl, 721, L82

\bibitem[Cavagnolo et al.(2009)]{2009ApJS..182...12C} Cavagnolo, K.~W., Donahue, M., Voit, G.~M., \& Sun, M.\ 2009, \apjs, 182, 12 
\bibitem[Clarke \& Ensslin(2006)]{2006AJ....131.2900C} Clarke, T.~E., \& Ensslin, T.~A.\ 2006, \aj, 131, 2900 
\bibitem[Condon et al.(1998)]{Condon98} Condon, J.~J., Cotton, W.~D., Greisen, E.~W., Yin, Q.~F., Perley, R.~A., Taylor, G.~B., \& Broderick, J.~J.\ 1998, \aj, 115, 1693 

\bibitem[Dallacasa et al.(2009)]{2009ApJ...699.1288D} Dallacasa, D., Brunetti, G., Giacintucci, S., et al.\ 2009, \apj, 699, 1288
\bibitem{} Dennison B., 1980, ApJ 239, L93
\bibitem[Dolag(2006)]{2006AN....327..575D} Dolag, K.\ 2006, Astronomische Nachrichten, 327, 575
\bibitem[Donnert et al.(2013)]{2013MNRAS.429.3564D} Donnert, J., Dolag, K., Brunetti, G., \& Cassano, R.\ 2013, \mnras, 429, 3564 

\bibitem{}Ebeling, H., Edge, A.C., B\"ohringer, H., et al.,1998, MNRAS, 301, 881
\bibitem{}Ebeling, H., Edge, A.C., Allen, S.W., et al., 2000, MNRAS, 318, 333
\bibitem[En\ss lin \& R\"ottgering (2002)]{ER02} En\ss lin T.A., R\"ottgering H., 2002, A\&A, 396, 83

\bibitem{}Feretti, L., 2002, in IAU Symp. 199, The Universe at Low Radio Frequencies, ed. A. Pramesh Rao, G. Swarup, \& Gopal-Krisna (San Francisco, CA: ASP), 133
\bibitem[Feretti(2003)]{F03} Feretti, L.\ 2003, Astronomical Society of the Pacific Conference Series, 301, 143
\bibitem[Feretti et al.(2012)]{2012A&ARv..20...54F} Feretti, L., Giovannini, G., Govoni, F., \& Murgia, M.\ 2012, \aapr, 20, 54 
\bibitem{}Ferrari, C., Govoni, F., Schindler, S., Bykov, A.M., Rephaeli, Y., 2008, Space Science Reviews, Vol. 134, p. 93

\bibitem[Giacintucci et al.(2009)]{2009A&A...505...45G} Giacintucci, S., Venturi, T., Brunetti, G., et al.\ 2009, \aap, 505, 45 
\bibitem[Giacintucci et al.(2011)]{2011A&A...534A..57G} Giacintucci, S., Dallacasa, D., Venturi, T., et al.\ 2011, \aap, 534, A57
\bibitem[Giacintucci et al.(2013)]{2013arXiv1302.0218G} Giacintucci, S., Kale, R., Wik, D.~R., Venturi, T., \& Markevitch, M.\ 2013, ApJ, 766, 18

\bibitem[Giocoli et al.(2007)]{2007MNRAS.376..977G} Giocoli, C., Moreno, J., Sheth, R.~K., \& Tormen, G.\ 2007, \mnras, 376, 977
\bibitem[Giocoli et al.(2012)]{2012MNRAS.422..185G} Giocoli, C., Tormen, G., \& Sheth, R.~K.\ 2012, \mnras, 422, 185

\bibitem[Giovannini et al.(1999)]{GTF99} Giovannini G., Tordi M., Feretti L., 1999, NewA 4, 141

\bibitem[Giovannini et al.(2009)]{2009A&A...507.1257G} Giovannini, G., Bonafede, A., Feretti, L., et al.\ 2009, \aap, 507, 1257


\bibitem{}Govoni, F., En\ss lin, T. A., Feretti, L., Giovannini, G., 2001, A\&A 369, 441
\bibitem{}Govoni, F., Markevitch, M., Vikhlinin, A., et al., 2004, ApJ, 605, 695
\bibitem[Govoni et al.(2005)]{2005A&A...430L...5G} Govoni, F., Murgia, M., Feretti, L., et al.\ 2005, \aap, 430, L5

\bibitem[Isobe et al.(1986)]{1986ApJ...306..490I} Isobe, T., Feigelson, E.~D., \& Nelson, P.~I.\ 1986, \apj, 306, 490
\bibitem[Isobe et al.(1990)]{1990ApJ...364..104I} Isobe, T., Feigelson,  E.~D., Akritas, M.~G., \& Babu, G.~J.\ 1990, \apj, 364, 104



\bibitem[Jeltema \& Profumo(2011)]{2011ApJ...728...53J} Jeltema, T.~E., \& Profumo, S.\ 2011, \apj, 728, 53 

\bibitem[Kale et al.(2013)]{2013arXiv1306.3102K} Kale, R., Venturi, T., Giacintucci, S., et al.\ 2013, arXiv:1306.3102 
\bibitem[Kempner \& Sarazin(2001)]{KS01} Kempner, J.~C., \& Sarazin, C.~L.\ 2001, \apj, 548, 639
\bibitem[Kim et al.(1990)]{1990ApJ...355...29K} Kim, K.-T., Kronberg, P.~P., Dewdney, P.~E., \& Landecker, T.~L.\ 1990, \apj, 355, 29
\bibitem[Krause et al.(2012)]{2012MNRAS.419.1766K} Krause, E., Pierpaoli, E., Dolag, K., \& Borgani, S.\ 2012, \mnras, 419, 1766
\bibitem[Kushnir et al.(2009)]{2009JCAP...09..024K} Kushnir, D., Katz, B., \& Waxman, E.\ 2009, JCAP, 9, 24 

\bibitem[Liang(1999)]{Liang99} Liang H., 1999, in "Diffuse thermal and relativistic plasma in galaxy clusters". Edited by Bohringer H., Feretti L., Schuecker P.. Garching, Germany : Max-Planck-Institut fur Extraterrestrische Physik, 1999. ("Proceedings of the Workshop...Ringberg Castle, Germany, April 19-23, 1999".), p.33
\bibitem{} Liang H., Hunstead R.W., Birkinshaw M., Andreani P., 2000, ApJ 544, 686



\bibitem[Macario et al.(2010)]{2010A&A...517A..43M} Macario, G., Venturi, T., Brunetti, G., et al.\ 2010, \aap, 517, A43 
\bibitem[Macario et al.(2011)]{2011ApJ...728...82M} Macario, G., Markevitch, M., Giacintucci, S., et al.\ 2011, \apj, 728, 82 

\bibitem[Markevitch(1998)]{1998ApJ...504...27M} Markevitch, M.\ 1998, \apj, 504, 27
\bibitem{}Markevitch, M.; Vikhlinin, A., 2001, ApJ 563, 95

\bibitem[Marrone et al.(2012)]{2012ApJ...754..119M} Marrone, D.~P., Smith, G.~P., Okabe, N., et al.\ 2012, \apj, 754, 119 
\bibitem{}Maughan, B. J.; Jones, C.; Forman, W.; Van Speybroeck, L.,2008, ApJS 174, 117
\bibitem{}Mohr, J.J., Fabricant, D.G., Geller, M.J., 1993, ApJ 413, 492
\bibitem[Motl et al.(2005)]{2005ApJ...623L..63M} Motl, P.~M., Hallman, E.~J., Burns, J.~O., \& Norman, M.~L.\ 2005, \apjl, 623, L63 

\bibitem[Nagai(2006)]{2006ApJ...650..538N} Nagai, D.\ 2006, \apj, 650, 538 

\bibitem{} O'Hara, T.B., Mohr, J.J., Bialek, J.J., Evrard, A.E., 2006, ApJ 639, 64O

\bibitem[Petrosian (2001)]{Pe01} Petrosian V., 2001, ApJ 557, 560 

\bibitem{}Poole, G.B., Fardal, M.A., Babul, A. et al., 2006, MNRAS 373, 881

\bibitem[Planck Collaboration et  al.(2011)]{2011A&A...536A...8P} Planck Collaboration, Ade, P.~A.~R., Aghanim, N., et al.\ 2011a, \aap, 536, A8

\bibitem[Planck Collaboration et 
al.(2011)]{2011A&A...536A..11P} Planck Collaboration, Ade, P.~A.~R., Aghanim, N., et al.\ 2011b, \aap, 536, A11 

\bibitem[Planck Collaboration et al.(2011)]{2011A&A...536A...9P} Planck Collaboration, Aghanim, N., Arnaud, M., et al.\ 2011c, \aap, 536, A9

\bibitem[Planck Collaboration et al.(2013)]{} Planck Collaboration, Ade, P.~A.~R., Aghanim, N., et al.\ 2013a, A\&A, 554, A140

\bibitem[Planck Collaboration et al.(2013)]{2013arXiv1303.5089P} Planck Collaboration, Ade, P.~A.~R., Aghanim, N., et al.\ 2013b, arXiv:1303.5089 

\bibitem[Pratt et al.(2009)]{2009A&A...498..361P} Pratt, G.~W., Croston, J.~H., Arnaud, M., B{\"o}hringer, H.\ 2009, \aap, 498, 361

\bibitem[Reiprich]{2002ApJ...567..716R} Reiprich, T.~H., B{\"o}hringer, H.\ 2002, \apj, 567, 716

\bibitem[Rossetti et al.(2011)]{2011A&A...532A.123R} Rossetti, M., Eckert, D., Cavalleri, B.~M., et al.\ 2011, \aap, 532, A123

\bibitem[Rudnick et al.(2006)]{2006AN....327..549R} Rudnick, L., Delain,  K.~M., \& Lemmerman, J.~A.\ 2006, Astronomische Nachrichten, 327, 549 
\bibitem[Rudnick \& Lemmerman(2009)]{2009ApJ...697.1341R} Rudnick, L., \& Lemmerman, J.~A.\ 2009, \apj, 697, 1341 

\bibitem[Russell et al.(2011)]{2011MNRAS.417L...1R} Russell, H.~R., van Weeren, R.~J., Edge, A.~C., et al.\ 2011, \mnras, 417, L1 
\bibitem[Russell et al.(2013)]{2013MNRAS.432..530R} Russell, H.~R., McNamara, B.~R., Edge, A.~C., et al.\ 2013, \mnras, 432, 530

\bibitem{}Santos, J.S.; Rosati, P.; Tozzi, P.; et al., 2008, A\&A, 483, 35

\bibitem[Schlickeiser et al.(1987)]{Sch87} Schlickeiser R., Sievers A., Thiemann H., 1987, A\&A 182, 21
\bibitem{}Schuecker, P., B\"ohringer, H., Reiprich, T.H., Feretti, L., 2001,  A\&A, 378, 408


\bibitem[van Weeren et al.(2009)]{2009A&A...505..991V} van Weeren, R.~J., R{\"o}ttgering, H.~J.~A., Br{\"u}ggen, M., \& Cohen, A.\ 2009, \aap, 505, 991 
\bibitem{}Ventimiglia, D.A., Voit, G.M., Donahue, M., Ameglio, S., 2008, ApJ 685, 118

\bibitem{}Venturi, T., Giacintucci, S., Brunetti, G., et al., 2007, A\&A 463, 937
\bibitem{}Venturi, T., Giacintucci, S., Dallacasa, D., et al., 2008, A\&A, 484, 327
\bibitem[Venturi et al.(2013)]{2013A&A...551A..24V} Venturi, T., Giacintucci, S., Dallacasa, D., et al.\ 2013, \aap, 551, A24 

\bibitem[Zhang et al.(2007)]{2007A&A...467..437Z} Zhang, Y.-Y., Finoguenov, A., B{\"o}hringer, H., et al.\ 2007, \aap, 467, 437

\end{thebibliography}
\end{document}